\documentclass{article}
\usepackage[onecolumn]{emulateapj}
\usepackage{epsf}

\def\upm{\hbox{$.\!\!^{\rm m}$}}
\def\ups{\hbox{$.\!\!^{\rm s}$}}
\def\updg{\hbox{$.\!\!^\circ$}}

\def\uparcs{\hbox{$.\!\!^{\prime\prime}$}}

\begin{document}

\submitted{Accepted for Publication in the Astrophysical Journal}

\title{Dynamics of the Globular Cluster System 
Associated with M87 (NGC 4486). 
I. New CFHT MOS Spectroscopy and the Composite Database}

\author{David A. Hanes\altaffilmark{1,2,3,4}, Patrick 
C\^ot\'e\altaffilmark{1,5,6,7}, Terry J. Bridges\altaffilmark{1,3,8,9},
Dean E. McLaughlin\altaffilmark{1,10,11}, Doug Geisler\altaffilmark{12,
13,14}, Gretchen L.H. Harris\altaffilmark{15}, James E. Hesser\altaffilmark{4},
Myung Gyoon Lee\altaffilmark{16}}

\altaffiltext{1}{Visiting Astronomer, Canada-France-Hawaii Telescope, operated
by the National Research Council of Canada, the Centre National de la Recherche
Scientifique of France, and the University of Hawaii}

\altaffiltext{2}{Department of Physics, Queen's University, Kingston ON  Canada K7L 3N6}

\altaffiltext{3}{Anglo-Australian Observatory, P.O. Box 296, Epping NSW 1710,
Australia}

\altaffiltext{4}{Dominion Astrophysical Observatory, Herzberg Institute of Astrophysics,
National Research Council, 5071 West Saanich Road, Victoria BC V9E 2E7, Canada}

\altaffiltext{5}{Department of Physics and Astronomy, Rutgers University, New Brunswick,
NJ 08854, USA}

\altaffiltext{6}{California Institute of Technology, Mail Stop 105-24, 
Pasadena CA 91125  USA}

\altaffiltext{7}{Sherman M. Fairchild Fellow}

\altaffiltext{8}{Royal Greenwich Observatory, Madingley Road, Cambridge, CB3 0EZ,
UK}

\altaffiltext{9}{Institute of Astronomy, Madingley Road, Cambridge, CB3 0HA, UK}

\altaffiltext{10}{Department of Astronomy, University of California, 
601 Campbell Hall, Berkeley CA 94720-3411  USA}

\altaffiltext{11}{Hubble Fellow}

\altaffiltext{12}{Grupo de Astronom\a'{\i}a,  Dpto. de F\a'{\i}sica,
Universidad de Concepci\'on, Casilla 160-C,
Concepci\'on, Chile}

\altaffiltext{13}{Visiting Astronomer, Cerro Tololo Inter-American
Observatory, which is operated by AURA, Inc., under cooperative agreement
with the National Science Foundation}

\altaffiltext{14}{Visiting Astronomer, Kitt Peak National 
Observatory, which is operated by AURA, Inc., under cooperative agreement
with the National Science Foundation}

\altaffiltext{15}{Department of Physics, University of Waterloo, 
Waterloo ON  Canada  N2L 3G1}

\altaffiltext{16}{Astronomy Program, SEES, Seoul National University, Seoul
151-742, Korea}

\lefthead{{\sc HANES ET AL.}}
\righthead{DYNAMICS OF THE M87 GLOBULAR CLUSTER SYSTEM}

\begin{abstract}

We present a comprehensive database of kinematic, photometric and positional
information for 	352 objects in the field of M87 (NGC 4486), the central giant
elliptical galaxy in the Virgo cluster; the majority of the tracers are globular
clusters associated with that galaxy.  New kinematic information comes from 
multi-slit observations with the Multi-Object Spectrograph (MOS) of the Canada-France-Hawaii
Telescope (CFHT), an investigation which has added 96 new velocities to and confirmed many of the
earlier values in a pre-existing dataset
of 256 velocities published elsewhere. 
 The photometry, consisting of magnitudes and
colors in the Washington (T$_1$, C-T$_1$) system, is based on CCD observations made 
at the Cerro Tololo Inter-American Observatory (CTIO) and the Kitt Peak National
Observatory (KPNO).  The composite database represents the largest compilation of pure 
Population II dynamical tracers yet identified in any external galaxy; moreover, it extends to larger
spatial scales than have earlier investigations. The inclusion of photometric 
information allows independent study of the distinct red and blue sub-populations of the
bimodal GCS of M87.  In a companion paper (C\^ot\'e et al. 2001), we
use this powerful dataset to analyse the present dynamical state of the M87 globular cluster system, 
and consider the question of its interaction and formation history.

\end{abstract}


\section{Introduction}

The use of globular clusters as dynamical tracers of galaxy halos, both within and
outside the Milky Way, has a long and
important history. Their importance transcends
that of mere descriptors of the present mass distribution and dynamical state of
the parent galaxy: their extreme ages and ranges of chemical composition, and 
the evident correlations between these attributes and cluster position and kinematics,
make them powerful fossil tracers of the  formation and interaction
history of galaxies.  The concept of a monolithic galaxy formation
process (Eggen, Lynden-Bell \& Sandage 1962), with attendant dissipation and chemical 
enrichment, has been supplanted in
recent years through the recognition of evidently more complex histories.  The
discovery of bimodal color distributions in the globular
cluster systems (GCSs) of many
elliptical galaxies, for instance, has invigorated theoretical investigations into the
importance of (inter alia) merger events (Ashman \& Zepf 1992), multi-modal star formation
histories (Harris, Harris \& McLaughlin 1998) and hierarchical growth 
(C\^ot\'e, Marzke \& West 1998) in 
the galaxy formation process.  These issues are explored in more depth in 
our companion paper (C\^ot\'e et al. 2001).
 
The promise of spectroscopic studies of GCSs is nowhere more evident than in external galaxies, 
within which globular clusters may serve 
as luminous dynamical test particles at galactocentric radii where the halo surface
brightness is impractically faint. Moreover, they are sufficiently numerous that reasonably
robust statistical analyses can be carried out, at least in principle.  
Until recently, however, heroic efforts were required to establish even limited datasets:
the elliptical galaxies known to possess large GCSs (Hanes 1977) typically lie at Virgo-like
distances, and spectroscopic investigations of their faint (V$\ge19$ mag) globular
clusters were at the limits of the available instrumentation; see for instance
Mould et al. (1990).

The development of multi-spectroscopic instruments, such as the MOS at the
Canada-France-Hawaii Telescope, has made feasible studies once beyond the grasp of
4-metre class telescopes: these instruments provide not just a considerable multiplex advantage,
but also the benefits of highly efficient optical throughputs at superb observing sites.
Such considerations motivated the extensive studies reported in Cohen and Ryzhov (1997) 
and Cohen (2000),
who investigated the M87 GCS using the LRIS multi-slit spectroscopic 
instrument at the Keck Telescope.  The data provided therein have been variously interpreted 
(Cohen and Ryzhov 1997; Kissler-Patig and Gebhardt 1998; Cohen, Ryzhov and Blakeslee 1998), but
suffer both from the somewhat limited spatial coverage and a lack of color
information which would allow independent analyses of the GCS sub-populations.  In the 
present paper, we report the results of observations carried out at the CFHT, one in which
we have been able to extend the spatial coverage, amplify the dataset by some 40\% in 
total size, and provide the photometric
information which will permit the comprehensive analysis reported in 
C\^ot\'e et al. (2001).

\section{Target Identification and Photometry}

The globular cluster system associated with M87 has been
the subject of a number of photometric investigations.  
In several of these (Hanes 1977; Strom et al. 1981; Harris 1986;
McLaughlin, Harris and Hanes 1994; Geisler, Lee and Kim 2001) the
emphasis has been upon the determination and analysis of the
global photometric properties, cluster
luminosity function, or spatial distribution of the system.
Recent HST studies (e.g., Whitmore et al. 1995; Elson and Santiago 1996;
Kundu et al. 1999) have mainly
focussed on the {\sl central} properties of the M87 globular cluster system (GCS).
For instance, from the V-I colors of roughly one thousand globular clusters in
the center of M87, Whitmore et al. (1995) and Elson and Santiago (1996) were the 
first to present  
unambiguous evidence for a bimodal color distribution -- a ubiquitous
feature among giant elliptical galaxies, and one which has important
implications for galaxy formation models.

Indeed, the principal aim of the present study is to create a database
which permits a subsequent dynamical analysis of the system and its
sub-components, whether selected by color or position, in analyses
intended to address the question of M87's formation history (C\^ot\'e et al. 2001).
Clearly, unambiguous target positions and precise multi-color 
photometry are necessary if this goal is to be realized.
In this section, we describe the foundation on which we have built 
our final optimal dataset.

\subsection{Strom et al (1981)}

Strom et al. (1981), hereinafter S81, analysed 
microdensitometer scans of a set of 
deep KPNO 4-metre prime focus plates in three colors (UBR); they
identified  1728 candidate globular clusters associated with M87.  
In subsequent spectroscopic studies 
(Mould, Oke and Nemec 1987; Huchra and Brodie 1987; Mould et al. 1990;
Cohen and Ryzhov 1997; Cohen, Blakeslee and Ryzhov 1998; Cohen 2000),
the practice has been to
identify targets by their S81 numbers, a convention
which we will continue where possible (but see
Section~\ref{Revised_No}).  

The S81 tabulations, although extensive, are 
inconvenient in one respect.  
Their X and Y positional entries 
(which we will hereafter denote with the symbols SX and SY) are 
in units of arcseconds relative to a
reference point near the SE plate corner; 
as a consequence, they lack an absolute 
astrometric calibration. 
To remedy this, we have carried out an 
astrometric calibration of the S81 
tabulations:
to an rms precision of about 
one arcsecond in each coordinate, the SX 
and SY coordinates tabulated by S81 are related to 
right ascension and declination according to:

\begin{equation}
		R.A.(2000)  =  12^{h}  +  30^{m} 
	+  ((1187.7  -  0.0244 SX - 1.0242 SY) / 15)^{s},
\end{equation}

\begin{equation}
		\delta(2000)  =   12^{o}  +  (994.8  +  
	1.0007 SX 
 -  0.0249 SY)^{"}.
\end{equation}

\noindent 
The M87 galaxy center itself, which is  located at $RA =  12^{h}  30^{m}  
49\ups4, \delta =  +12^{o}  23^{'}  
28^{"} (2000),$  lies at (SX,SY) = (424,426) in the S81 system. 
The SY term in Equation 1 can be recognized as 
$1.000/cos(\delta)$,
with $\delta \sim 12^{o} 23^{'}$; along with the SX term in Equation 2, this
confirms the precise correctness of the plate scale adopted by S81.  The small cross
terms indicate that the S81 coordinates were slightly rotated (by $\sim 1\updg4$) relative to the
cardinal directions on the sky.

In Table 1, we present a summary tabulation of the positional 
data for all the targets in the
M87 field for which spectroscopically derived velocities are now 
known, whatever the source.  (We postpone a discussion of the photometric and 
kinematic data in the table until Section~\ref{Phot_Tables} and Section~\ref{Velocities}.)
In Table 1, the first column represents the
S81 running number, amended  when necessary (or newly defined) for the reasons
described in Section~\ref{Revised_No}.  Subsequent columns present
the (SX,SY) coordinates (columns 2,3); the astrometrically-determined 
target right ascension and declination (columns 4-5); and the offsets with respect 
to the M87 nucleus, measured in seconds of arc positively to the East 
(column 6) and the North
(column 7).

\subsection{Geisler, Lee and Kim 2001} \label{GL_Phot}

In C\^ot\'e et al. (2001),
we  analyse  the dynamical behaviour 
of the red and blue sub-populations of the M87 globular cluster system. 
Such an analysis requires 
not just photometry of the several hundred dynamical tracers for which we now
have velocities, but also a broader understanding of the 
separate {\em surface density profiles} of the two sub-populations. 
 Unfortunately, the S81 photographic 
dataset is neither precise enough in its photographically-determined
colors nor deep enough to define sufficiently large 
samples for this purpose.
Indeed, many of  the relatively
bright spectroscopic targets themselves 
lack the precise colors necessary for their confident classification 
into the `red' and `blue' sub-populations.
Moreover, there are occasional
ambiguities in the target list; see Section~\ref{Revised_No}. 

The necessary information comes, however, from Geisler, Lee and Kim (2001),
who have carried out deep photometry in the Washington photometric 
system in the field of M87.  Their 
dataset was produced from direct CCD images taken at the prime 
focus of the KPNO 4-metre telescope.  
Using this extensive tabulation, we
have cross-identified as many as possible of 
our spectroscopic targets and those previously studied by Cohen and Ryzhov (1997),
Cohen (2000), and Mould et al. (1990),  extracting the T$_1$ magnitudes and C-T$_1$ 
colors which are shown in our 
final composite tabulation (see Section~\ref{Phot_Tables}).  
(It is important to note that the photometric data in the pre-publication 
tabulations provided by Geisler, Lee and Kim (2001)
may yet be subject to minor revisions of individual data points.
We do not expect such revisions to materially
affect any of the analysis described in C\^ot\'e et al. (2001).)

Even with the inclusion of the Geisler, Lee and Kim (2001)
data, however, a number of our
spectroscopic targets lacked essential 
color information, primarily because the direct images 
of Geisler, Lee and Kim (2001) were slightly offset from the center of M87.
As a consequence, a few dozen 
objects for which we obtained good 
spectra lay beyond the bounds of the photometric dataset.  
To make up for this deficiency, 
we acquired additional 
direct frames of the field at the prime focus of the 
CTIO 4-metre telescope in March 2000, using C and R 
filters in front of the CCD mosaic camera. 
Photometric calibration was accomplished through the study of 
objects common to both studies; in this way,  
the (R, C-R) magnitudes and colors were reliably reduced to 
the Washington (T$_1$, C-T$_1$) system.  In the end,
we were able to determine the necessary colors and 
magnitudes for essentially all of the objects 
studied spectroscopically, with a few exceptions to be discussed 
later.  At the relatively bright magnitude levels for objects 
for which we have reliable velocities, the photometric precision
is typically $\pm 0\upm02$ or better in both T$_1$ and (C-T$_1$).

\begin{figure*}[t]
\centering \leavevmode
\epsfysize=3.5truein
\epsfbox{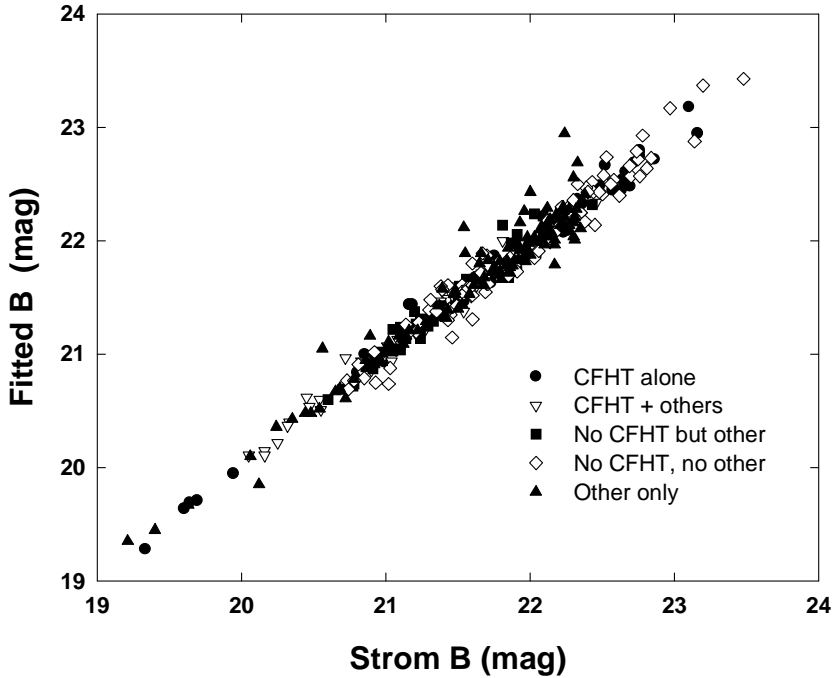}
\caption{A comparison of the (C,C-T$_1$) CCD photometry with the S81 photographic
B photometry, using the transformation given in the text, equation (3).  The symbols
represent: objects with CFHT velocities but no others reported (filled circles); objects with
velocities from CFHT and elsewhere (open triangles); objects unsuccessfully targeted at CFHT but with
velocities from elsewhere (filled squares); objects unsuccessfully targeted at CFHT and with no
published velocity (open diamonds); and objects not studied at CFHT but with published velocities
from other studies (filled triangles).
\label{fig1}}
\end{figure*}

As Geisler (1996) demonstrated, reliable transformations can be
established from the Washington (T$_1$, C-T$_1$)
system to UBVRI with an rms scatter of about 0.05 mag in 
photoelectric determinations. 
In Figure 1, we compare the B photographic photometry tabulated by
S81 to that derived from the CCD (T$_1$, C-T$_1$) study,  
with the datapoints restricted to those for which there is 
absolutely no ambiguity about
target identification (see Section~\ref{Revised_No}).  
A straightforward
least-squares analysis yields a best-fit given by:

\begin{equation}
  B_{fitted} = 1.069 (C) - 0.419 (C-T1) - 1.038.
\end{equation}

Although the leading coefficient reveals that there
is a small scale error in the photographic photometry, the tightness of
 the relationship (about which there is an rms scatter of $\sim 0.1$ mag) is a testament to
the quality of the photographic photometry presented by S81. 
(The various symbols are explained in the figure caption.)
Almost without exception, the filled triangles which lie above the mean relationship
represent targets which are 
projected against the central parts of M87, 
where the background is both bright and rapidly spatially variable.

\subsection{A Revised Numbering Scheme} \label{Revised_No}

Extensions to the S81
numbering scheme were introduced by Cohen and Ryzhov (1997) and
Cohen (2000) to permit the inclusion of targets not
appearing in the original photographic study. 
We need now to further extend  the S81 numbering scheme, for two reasons.  The first
is that the study 
reported here extends to spatial scales beyond those of S81.
But we have also discovered, through close inspection of our direct frames,
 that some of the S81 identifications do not
correspond to unique targets.
The tabulated coordinates for Strom 978, for instance, point to a
location in the middle of a very close triplet of objects, apparently 
merged in the S81 photographic analysis.  

Our objective is to provide a definitive database for subsequent analysis.
For absolute clarity, therefore, 
we have amended the numbering scheme in ways which accommodate
new objects, reflect the status of ambiguous targets, and indicate the
availability and reliability of 
 the necessary photometry.  In Table 2, we present a succinct target-by-target summary of the
considerations which motivate the amendments.  

{\em For all objects with published velocities,} we now adopt a set of identification
numbers which adhere to the following conventions:

{\bf Targets numbered below 2000} (289 of them): are those which retain 
their original S81 numbers to indicate that
they are
unambiguously identified and that we have 
good photometry in the (T$_1$,C-T$_1$) system.

{\bf Targets numbered in the low 5000s} (7 of them): were introduced by 
Cohen and Ryzhov (1997) to represent objects not found in S81 (typically because they
were rather close to the galaxy center and thus projected onto a bright 
background). Velocities were reported for seventeen such objects, although the
target numbers ranged from 5001 to 5028; presumably the missing ones refer to
slitlet positions  which yielded no useful spectra.
In Cohen and Ryzhov (1997), the positions of these objects were given
in terms of EW and NS offsets relative to Strom 928.  

We have found an 
unambiguous and precise $(\le 0\uparcs3)$ coordinate transformation which 
maps nine of these 
objects, along with Strom 928 itself, into bright targets in the 
Geisler, Lee and Kim (2001) photometric database,
and are completely confident of their identifications.  Of these nine objects,
seven retain the identification numbers introduced by Cohen and Ryzhov (1997) to reflect this
complete knowledge. 

 Two of  the identification numbers, 5024 and 5026,  are redundant.
It appears that the duplication of a couple of slit positions passed unnoticed in the 
analysis of the masks used at the telescope by Cohen and Ryzhov (1997).
Our astrometric analysis reveals that 
object 5026 corresponds 
to Strom 978 in Table 3 of Cohen and Ryzhov (1997), while object 5024 corresponds to the entry
immediately following Strom 978, an unnumbered object described in
a footnote as lying $6\uparcs6$ to the East.
This interpretation  is confirmed by the
good agreement in the velocities: Cohen and Ryzhov (1997) report
$v = 1076$ and $1141$ km/sec 
for objects 5024 and the unnamed target; meanwhile, they find $v = 1990$ and $1878$ km/sec for 
objects 5026 and 978.  For each pair, the measurements agree to well within the combined 
measurement errors 
estimated by Cohen and Ryzhov (1997), and we adopt unweighted average values in our
final tabulation.  The redundancy is complicated by the fact that
the targets lie at the notional position of 
object 978, which is itself a close triplet of unresolved objects.  (Indeed,
Cohen and Ryzhov (1997)
discovered the extra spectrum of their unnamed object serendipitously on the slit positioned 
on the other target.) 
We resolve the redundancy and reflect the ambiguity by renaming these objects as numbers
8005 (=5026/978)
and 8006 (=5024/unnamed), a choice explained below.

{\bf Targets numbered in the upper 5000s} (8 of them): represent the remaining 
``5000-series'' targets introduced by Cohen and Ryzhov (1997), those for which we could
find no counterpart in the Geisler, Lee and Kim (2001) data (although they appear on our
direct images).  All but one of these lie considerably
closer to the galaxy center, and thus on much brighter 
backgrounds, than do those for which 
we have made secure identifications.  To reflect the lack of photometric 
information, we renumber them in the upper 
5000s through the simple expedient of adding 50 to the numbers used by Cohen and Ryzhov (1997).
Although such targets can be used in undifferentiated analyses of the global
dynamical state of the cluster system, 
they can of course not be assigned to one of the cluster sub-populations
until new photometry is secured.

{\bf Targets numbered in the 6000s} (2 of them): represent two new objects presented by Cohen (2000),
with positions expressed in terms of offsets relative to
Strom 176.  Again we find secure identifications in the Geisler, Lee and Kim (2001) photometric
dataset, and simply retain the numbers introduced by Cohen (2000).

{\bf Targets numbered in the 7000s} (28 of them): represent, in a logical extension of the
numbering scheme introduced by Cohen and Ryzhov (1997) and Cohen (2000), objects not
appearing in the S81 tabulation but for
which we have good photometry and reliable CFHT MOS velocities and positions.
The inclusion of such objects stems from the fact that 
our mosaic of spectroscopic target fields 
extended beyond the boundaries of the S81 study.  

{\bf Targets numbered from 8001 to 8008} (8 of them): represent objects
which are 
found in the S81 dataset, but close inspection of which reveals that there
are potential ambiguities in identification. Typically, such objects 
consist of small clumps of several unresolved sources which the S81
tabulations do not distinguish.  Inspection of the MOS mask reveals which
component was the precise spectroscopic target, and 
they are 
likewise cleanly resolved in the Geisler, Lee and Kim (2001) photometry.
Such targets, therefore, can be used without qualm in the ensemble dataset.

{\bf Targets numbered from 8051 to 8056} (6 of them): represent objects which are also 
ambiguous in the S81 dataset  but for which we
can still confidently assign a photometric classification into the `blue' or `red'
sub-population.  For instance, Cohen and Ryzhov (1997) report a velocity for Strom 827 
but did not record that the (SX,SY) coordinates in fact point to a close {\em pair} of
unresolved objects. Without reference to the LRIS masks, we cannot know which
of them was the spectroscopic target, but the Geisler, Lee and Kim (2001) photometry yields the
information that both components are clearly red (according to the criterion
adopted in C\^ot\'e et al. (2001)). Except for a very small uncertainty in the 
precise position, therefore, such targets can be used with confidence in the 
multi-color dynamical analysis described in the companion paper C\^ot\'e et al. (2001).

{\bf Targets numbered in the low 9000s} (2 of them): represent objects which are ambiguous in
the S81 data set and for which we are unable to carry out a photometric 
classification.  By contrast to the objects in the `high-8000' series, these
are typically double targets unresolved by S81 for which
Keck velocities are reported by Cohen and Ryzhov (1997) or Cohen (2000) but for which
there is a red and a blue component. Again, lacking access to the LRIS masks we
are unable to determine which component provided the spectrum.  Such tracers can
be used in a global dynamical analysis (modulo a very small positional ambiguity)
but cannot be use in the treatments of sub-populations identified by color. 

{\bf Targets numbered in the high 9000s} (3 of them):  have no reliable photometry.  One of these,
number 9052, formerly Strom 868,
is lost in an apparent cosmic ray event in the Geisler, Lee and Kim (2001) database; another, 
number 9051, is easily seen in our CFHT direct frames but not found in the
Geisler, Lee and Kim (2001) tabulations. The third object, number 9053 (formerly Strom 881), lay outside
the Geisler, Lee and Kim (2001) study, but unfortunately fell too near
the edge of the additional direct frames acquired at CTIO for reliable photometry. 
 Again, the velocities make these targets useful for
a global analysis but not in any treatment which entails a split into sub-populations
defined by color.

\subsection{The Final Dataset: Photometry and Kinematics} \label{Phot_Tables}

In  Table 1, in addition to the astrometric data, we present the composite 
photometric and 
kinematic dataset for 
all objects for which velocities
are now known (but defer a discussion of the velocity determinations until the 
next section).    Successive
columns provide 
the magnitudes and colors in the Washington T$_1$, (C-T$_1$) system (columns 8,9);
the new CFHT MOS-based velocities and their
associated formal uncertainties (column 10); and previously published velocities
from studies at the Keck telescope (K00 = Cohen (2000); K97 = Cohen and Ryzhov (1997)) and elsewhere 
(M90 = Mould et al. (1990)) (columns 11-13).

\section{CFHT MOS Spectroscopy}

\subsection{Instrumentation and Methodology}

We used the MOS (Multi-Object Spectrograph) on five dark nights (May 17-22 1996) at 
the f/8 Cassegrain 
focus of the Canada-France-Hawaii Telescope (CFHT).   (An earlier scheduled run, 
in March 1995, was lost 
completely to a combination of island-wide power outages and bad weather.) 
The MOS instrument has been 
well described elsewhere (Le Fevre et al. 1994); in direct imaging mode, it consists of an 
f/8-f/2.75 focal-reducer which is 
used to reimage the Cassegrain field through transfer optics onto a 
2048x2048 CCD.  

The STIS2 CCD which is 
used with MOS has 21 micron square pixels, yielding an image scale 
of 0.44 arcsec per pixel and full-field 
coverage of 15 x 15 arcmin, although the MOS optics and mask
 holders introduce a vignetting which 
reduces this to an effective size of about 9.5 arcmin square
 in direct imaging mode.  (See Table 3 for a brief 
summary of these and other important points.)
Once acquired, the direct images are used in the design of masks containing 
slitlets overlying the desired 
spectroscopic targets.
  The entire process of designing, laser-cutting and 
inserting a mask into the four-position mask slide, and then aligning it to the 
target field, takes 
typically about an hour and a half. 

Early on the first night of our run, we 
had difficulty in securing well-focussed images for
mask design because of problems with the auto-focus software. 
Because of the 
less than optimal scheduling 
of the run (in late May), this meant that our principal field had set on Night 1
before we had designed, cut, and mounted the first masks.  Thereafter, the operation
ran smoothly, except that on Night 4 one of our masks was accidentally inserted
into the holder slightly askew.  The targets on only about half of the field lay on the
slitlets, but still provided useful spectra; for the last night's observations, 
we cut and mounted a new mask. 

In designing the masks, we identified our targets first with reference to the V band data of
McLaughlin, Harris and Hanes (1994);
in more distant regions of the GCS, where no published
CCD photometry exists, targets were selected solely on the basis of 
having a starlike appearance.  
Thanks to the unavailability of color information, 
our selection of targets necessarily included 
some which were unlikely to be globular clusters.
(Indeed, the objects of extreme red color typically turn out to have M-dwarf
spectra, as Cohen and Ryzhov (1997) also noted.)  The value of having a pre-existing
photometric dataset in multiplex investigations of this sort 
can scarcely be overstated. 
We will return to this point in our discussion of our attained efficiency. 

 For the dispersing 
element, we employed the B400 grism, which has a nominal dispersion of 
 $3.6\AA$ per
21-micron pixel and a 
zero-deviation wavelength of $5186\AA$, but inserted a pre-filter with a 
throughput from $4400\AA$ to $5700\AA$.  
This strategy allowed us to include extra targets in the
 CCD field without overlapping any spectra.  (Our 
most crowded field, for instance, had 104 slitlets.)  In general, slitlets were 
designed to a length of eight to ten arcsec and were  centrally placed on 
the targets; 
 the local sky was thus well sampled on
$\sim5-6$ arcsec scales spanning most of the objects.
 Here and there,
a number of somewhat shorter slitlets were  used
to allow the inclusion of additional targets.

\begin{figure*}[t]
\centering \leavevmode
\epsfysize=3.5truein
\epsfbox{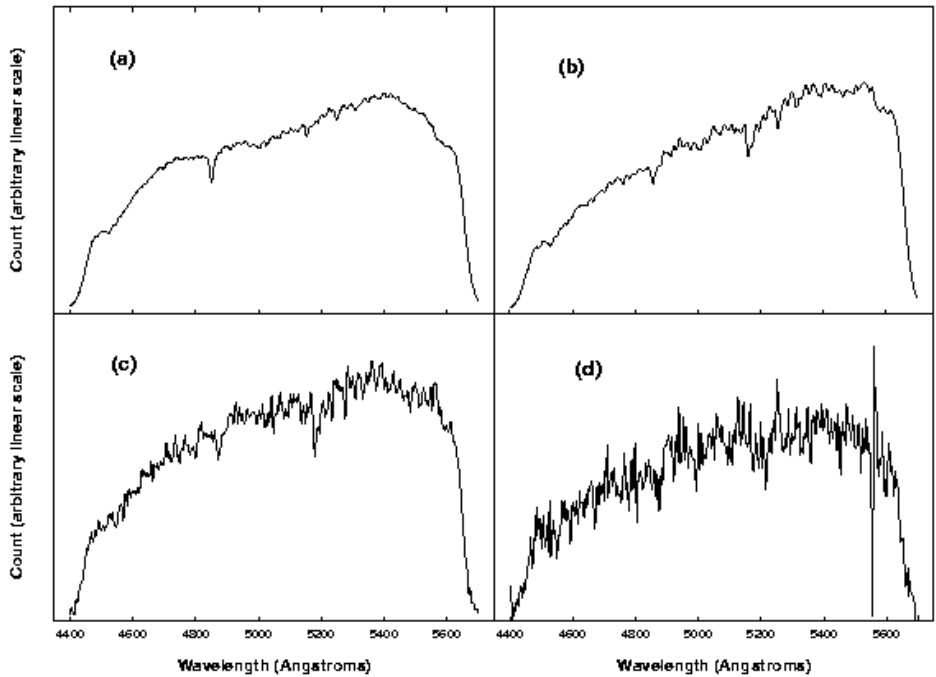}
\caption{Reduced spectra for (a) NGC 6205 (M13), a Galactic globular cluster; (b) NGC 6356, a
Galactic globular cluster; (c) Object 8007 in the M87 field, a target with 
(T$_1$, C-T$_1$) = (19.46, 1.58); and (d) Object 859 in the M87 field, a target with 
(T$_1$, C-T$_1$) = (20.98, 1.30).  The derived heliocentric velocities for objects 
8007 and 859 are 1183 $\pm$ 32 km sec$^{-1}$ and 1064 $\pm$ 91 km sec$^{-1}$ respectively.
\label{fig2}}
\end{figure*}

In Figure 2, we present a montage of
spectra for several objects: (a) the metal-poor Galactic globular cluster 
NGC 6205 (M13), for which [Fe/H]=$-1.54$; (b) the metal-rich Galactic globular
cluster NGC 6356, with [Fe/H]=$-0.50$; (c) the M87 globular cluster Strom 8007,
with (T$_1$,C-T$_1$)=(19.46,1.58); and (d) the M87 globular cluster Strom 859,
with (T$_1$,C-T$_1$)=(20.98,1.30).
The most prominent 
spectroscopic features 
are the Mgb triplet at $\sim5170\AA$ and $H\beta$ at 
$4861\AA$, although the 
cross-correlation techniques employed in the determination of 
velocities are of course sensitive to all 
features in the wavelength range.

\subsection{Instrument Characterization and Calibration}

The observing procedures were standard. 
In brief:

Approximately two dozen full bias frames were taken at various stages in 
the run.  Halogen (continuum) lamps were used to illuminate the 
masks and were imaged both with and without 
the grism in place---the latter to provide an unambiguous 
identification between any slit and its intended 
target in the direct image of the galaxy field, and the 
former to allow the tracing of any geometric 
(principally pincushion) distortion in the MOS optics. 
These continuum lamp images also permit the 
determination of the uniformity of the slit width along 
its length (the slit function). 
 
 Helium and neon emission lamps 
were used to provide wavelength calibration for the spectra acquired.   
Wavelength-calibration frames were taken frequently, 
and always before any movement of the telescope 
to a new target or change in position of the MOS mask
 holder.   During data reduction, the prominent night-sky line at
$\sim 5577\AA$ was monitored as an external check on our wavelength
calibration; its position was consistently as expected, typically to better than one-tenth of 
a pixel.

At opportune moments during the observing run, we acquired ten 
dark frames.
These frames took on an unforeseen 
 importance when we discovered that the 
MOS detector 
was registering a  diffuse light leak 
from an unknown source within the instrument rack.
Unfortunately, for the first three nights of the run  
this problem resisted a simple solution.
Although the light leak displayed 
 a degree of frame-to-frame 
variability,
 both in amplitude and in exact 
profile,
subtracting an interactively scaled version 
of the average dark frame compensated 
fairly well for the contaminating light,
leaving only small
residual gradients. 
 As will be shown, there are no indications of 
problems introduced as a result of this process: when we intercompare
our velocities to those from independent
studies, we find very good agreement. 
We are confident, therefore,
 that the our handling of the unwanted 
scattered light has introduced no significant systematic errors.

Inevitably, though, the subtraction of the unwanted light led to 
a degradation in the final 
spectroscopic signal-to-noise ratio in the affected fields.  
The effect of this is shown in Figure 3, in which we contrast the
efficiencies of our velocity determinations in frames secured 
with (panel (a)) and without (panel (b)) the light leak. 
In the figure, filled symbols represent targets for which 
reliable velocities were determined; open symbols represent targets
for which no good velocity could be derived.  (The differently shaped symbols
represent the different pointings of the telescope.)  In dark conditions,
the attained efficiency was almost exactly $50\%$; in the presence of the light leak, this
was dramatically reduced (to about $30\%$), with an obvious loss of limiting
magnitude.  The figure also demonstrates that there are also some rather bright
targets for which we failed to determine good velocities.  In many cases, this is 
because of the very poor template mismatch between our calibrating spectrum
(the globular cluster composite) and that of the target (often a recognizable  M-star
spectrum).

\begin{figure*}[t]
\centering \leavevmode
\epsfysize=3.5truein
\epsfbox{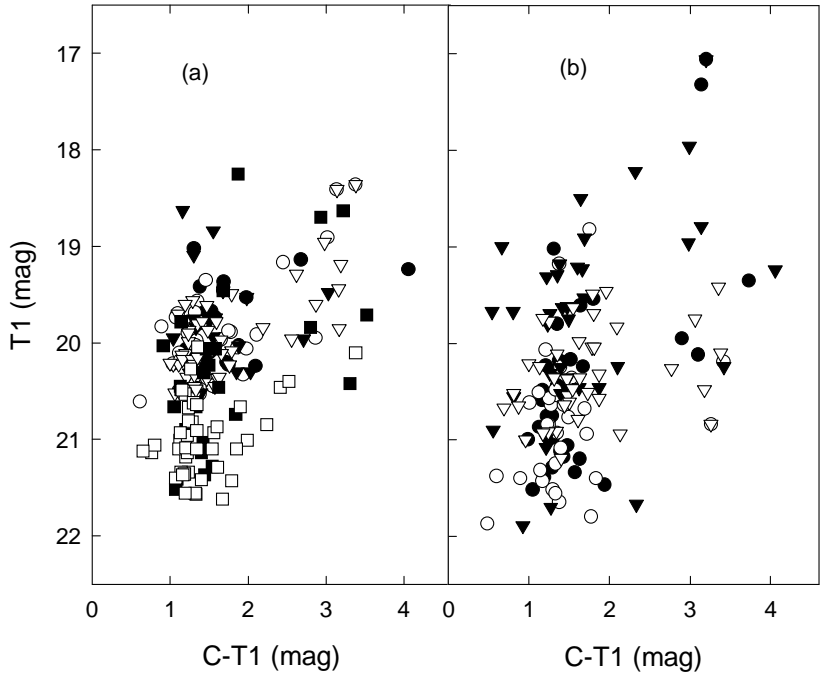}
\caption{Color-magnitude diagrams in (T$_1$,C-T$_1$) for targets in the M87 field 
for which good velocities
were derived (filled symbols) or not (open symbols).  Panel (a) corresponds to the three
early pointings for
which the observations were troubled by scattered light; panel (b) represents the two later
pointings in which this was  not a problem.  The loss of efficiency and the attendant
effect on our limiting magnitude are evident in panel (a).
\label{fig3}}
\end{figure*}

\subsection{Velocity Templates}

Spectra were acquired for radial velocity standard stars
of appropriate spectral type (G8III, K2III) and 
for a number of Galactic globular clusters (NGCs 6171, 
6205, 6356, 6402, 6528, 6624) spanning a range 
of metallicities.  For the globulars, the telescope was 
made to nod slowly during the integration in order 
to sweep a single long slit across the face of the target; 
after wavelength calibration and geometrical 
distortion corrections, the data were then further summed 
along the central parts of the slit to produce a 
final integrated spectrum.  The extreme ends of the slit 
were used to determine the contribution 
attributable to sky light.

In Table 4, we summarize the important properties 
of the Galactic globular clusters observed as radial velocity
templates.  Experimentation revealed that considerably more reliable
velocities resulted from their use (when cross-correlated one with 
another) than could be obtained when the bright standard stars were used as
velocity standards.
We attribute this to two factors: first, the Galactic globular 
clusters fill the slit uniformly, whereas the necessarily brief
integrations on the very bright radial velocity stars may  
introduce systematic offsets if the stars lie slightly off the precise slit center.
Secondly, the clusters have composite spectra, and are presumably 
better templates for one another -- and, by extension, for the target clusters in 
M87 -- than spectra of individual stars.

For these reasons, we elected to use only the Galactic globular clusters
as velocity references.  For each of them, an excellent high-signal-to-noise
spectrum was reduced to the rest frame, and the spectra were then normalized and added 
to provide a standard spectrum with composite features
spanning the range of likely globular cluster compositions.  The cross-correlation analysis
followed conventional practice; see Section~\ref{Reduction}.

\subsection{The M87 Mosaic} \label{Mosaic}

With the objective of studying the M87 GCS to large galactocentric 
distances, we secured a mosiac of 
five direct images on the first night: one central pointing; and 
four more offset radially by about 5.4 
arcmin in each of the NE, NW, SE, and SW directions (520 pixels, 
or 3.84 arcmin, along each of the 
cardinal directions).  Given the field size of the MOS detector, this
allowed overlap strips within which a number of targets were 
repeated; see Section~\ref{Internal}.
Moreover, the populous central field was visited twice,
using two different masks 
but with several of the targets repeated.
Unfortunately,
time lost to bad weather and occasional technical problems prevented 
us carrying out any spectroscopic 
observations in the NE quadrant on any of the scheduled  nights, but 
the other parts of the mosaic were 
well sampled. 
In Table 5, we present a summary description of the spectroscopic 
integrations carried out on the fields in 
our M87 mosaic. 

\subsection{Spectroscopic Data Reduction} \label{Reduction}

The data reduction was carried out almost entirely within 
the well-known suite of IRAF \footnote{IRAF is distributed by the National Optical
Astronomy Observatories, which are operated by the Association of
Universities for Research in Astronomy, Inc., under contract to the
National Science Foundation.} common-user software. The process was  
straightforward but necessarily quite interactive, with the following 
principal stages: 

Every frame was preprocessed in the usual way.  The overclocked 
regions were evaluated and 
subtracted; the median bias was removed; and an appropriately 
scaled `dark' frame was subtracted if 
necessary. 
Cosmic ray events were removed through a combination of the 
IRAF COSMICRAY task and special 
interactive software written expressly for the purpose.  

Individual spectra were identified in the data frames with 
reference to the corresponding direct images 
of the target fields and the illuminated masks. In the dispersed 
images, the typical slit length projected to less
than 30 pixels (13{$^{\prime\prime}$}), with the spectra stretching about 300 pixels 
in the sense of the dispersion.  Matched 
sub-frames of 40x320 pixels were removed from the spectroscopic 
frame, the emission lamp frame(s), 
and the dispersed halogen lamp (continuum) frame. 

For each triad of such frames, the continuum spectrum was 
examined to see if any geometrical 
transformation was needed (typically a small rotation) to 
remove the effects of pincushion distortion.  
When necessary, this correction was applied to all three of 
the excised sub-frames to straighten the 
spectra.

The continuum spectrum was collapsed spectrally over 
the regions of highest signal to yield a 
high-precision estimate of the slit function.  This was required 
for several of the spectra because of an 
occasional raggedness in the laser cutting of slitlets, 
a problem which has since been largely solved at 
the CFHT with the introduction of an auto-focus feature 
in the mask-cutting machinery.  
  For the vast majority of the spectra, 
however, the data needed at most a modest correction for 
 the slit function.

 The remaining steps in the reduction were conventional:
the definition of the location of the star along 
the slit; the determination of the level of and gradient 
in the sky (background) light; the wavelength 
calibration with reference to the corresponding arc frames;
 and so on.  Optimal weighting was used in 
the sky subtraction, and the final sky-subtracted target 
spectra were put onto linear wavelength scales, 
preserving the original spectral resolution. 

The final velocities and their respective uncertainties
were determined by use of the IRAF FXCOR package, 
and were derived independently 
by two of us to allow separate assessments of the 
strength and significance of the derived 
cross-correlation peaks and the influence of residual cosmic ray 
traces and the like.  Any objects which 
yielded significantly different velocities under these two 
treatments were given especially careful scrutiny; a 
few were rejected as untrustworthy.  
The median uncertainty of our CFHT/MOS velocities
for the 109 confirmed globular clusters in our sample (see 
Paper II) is ${103 \pm 20}$ km s$^{-1}$.

\subsection{The Velocity Precision Attained: Internal Comparisons} \label{Internal}

Before comparing our derived  velocities to those available from 
other studies, we carry out an 
intercomparison of those objects for which we derived reliable 
velocities from more than one mask.  
Unfortunately, there are not as many such targets as we had 
planned, thanks to the reduced success rates 
which we experienced on the masks afflicted by scattered
 light problems. Moreover, except for the 
repeated central pointings, the overlap regions were moderately 
close to the edges of the masks, in regions 
where slitlets were sometimes raggedly cut to such an extent 
that no useful spectrum could be derived even 
for a relatively bright object.  In the last analysis, however,
 we have seven multiply-studied objects, with every 
mask represented at least once.

\begin{figure*}[t]
\centering \leavevmode
\epsfysize=3.0truein
\epsfbox{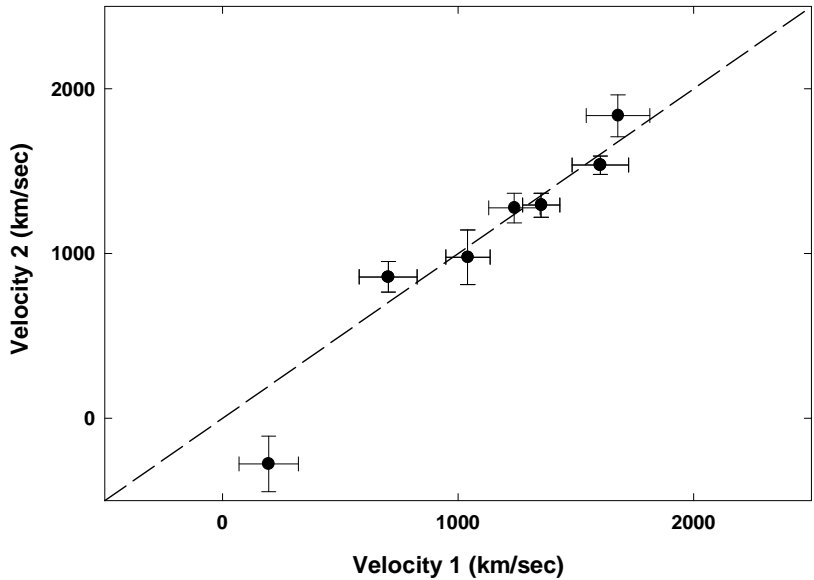}
\caption{A comparison of velocities derived for targets studied on independent 
MOS masks; the error bars represent the formal uncertainties.  In the various 
pairings, every one of the five masks employed is represented at least once. 
The broken line is the identity relationship.
\label{fig4}}
\end{figure*}

Figure 4 shows the quality of the systematic agreement.
It can be seen that the velocities agree to within 
the formal uncertainties, except for a single low-velocity object
for which the deviation from the identity line
is still less than twice the combined errors.  
This intercomparison reassures us that there are no 
significant mask-to-mask zero-point or scale differences, 
and that our estimates of the formal uncertainties are realistic.

\subsection{The Velocity Precision Attained: External Comparisons} 

Various studies (Huchra and Brodie 1987; Mould, Oke and Nemec 1987;
Mould et al. 1990; Cohen and Ryzhov 1997; Cohen 2000) have reported
velocities for 
the globular clusters associated with M87 (and inevitably for a number of the
field stars and compact 
background galaxies in the field).  The most comprehensive of these studies is that stemming 
from the use of LRIS at the
Keck Telescope,   the first 
tabulations of which appeared in Cohen and Ryzhov (1997).
These data were later augmented 
by the addition of five new datapoints (and two amendments) in Cohen, Blakeslee and Ryzhov (1998);
another sixteen velocities (four of which repeat values from Cohen and Ryzhov (1997))
appeared in Cohen (2000).
In total, these studies contribute 244 velocities for unresolved objects in the
M87 field,  
so it is particularly 
important to test the systematic agreement of our own dataset
with those measurements.

\begin{figure*}[b]
\centering \leavevmode
\epsfysize=3.5truein
\epsfbox{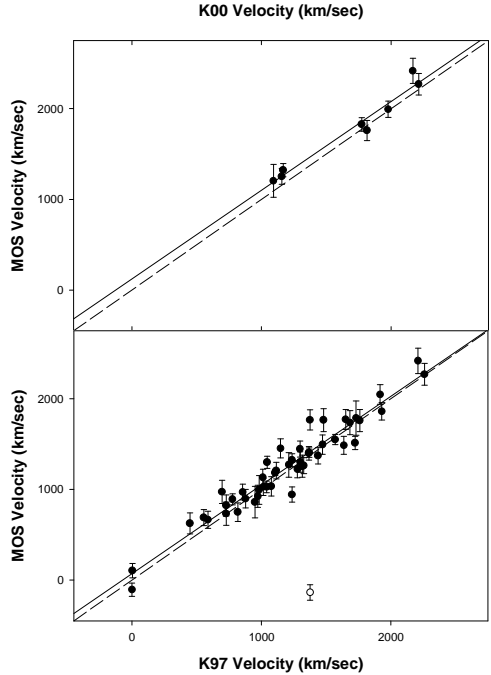}
\caption{A comparison of the velocities determined with the CFHT MOS instrument and 
thise from the studies of Cohen and Ryzhov (1997)
(bottom panel; 48 objects in common) and Cohen (2000) (top panel; 8 objects 
in common). The formal CFHT 
error bars are shown.  In each panel, the broken line represents the identity relationship, while the
solid line is the formal least-squares fit to the data points.  The open symbol in the bottom panel
corresponds to Object 321, which may have been misidentified in Cohen and Ryzhov (1997); the point was not
used in the regression analysis. [J.Cohen (2001, private communication) has now confirmed that the
LRIS spectrum of Object 321 is that of an M star of low velocity, consistent with the MOS result.]
\label{fig5}}
\end{figure*}

In Figure 5, therefore, we present a comparison of MOS-derived velocities, plotted
with explicit error bars, with those tabulated in Cohen (2000) (upper panel; 8 objects) and
Cohen and Ryzhov (1997) (lower panel; 48 objects).  In each case, the broken line represents the identity
relationship. The best-fit least-squares line, shown in solid, is not significantly
different from the line of unit slope in either case. In the lower panel in particular, 
the formal linear regression is given by
	\(v_{MOS} = 0.96\: (v_{K97}) + 91\: km\: sec^{-1}.\)

In the upper panel, there is a suggestion of a zero-point offset between
the MOS velocities and those reported by Cohen (2000), in the sense that the MOS
velocities are larger by $\sim$ 85 km sec$^{-1}$ on average; interestingly, for four
repeated objects, the velocities reported by Cohen and Ryzhov (1997) are likewise numerically
larger than those in Cohen (2000), 
by 62 km/sec in the mean, suggesting that  the zero-point in the Cohen (2000)
study may have an offset of about this magnitude relative to the Cohen and Ryzhov (1997) frame.

Although of much lower precision, the velocities tabulated by Mould et al. (1990) represent the
heroic efforts of various researchers Huchra and Brodie 1987; Mould, Oke and Nemec 1987) 
whose observations predated the advent of
multiplex devices such as MOS and LRIS.   Perhaps surprisingly, not all of the rather bright
targets which they studied have been repeated in subsequent programs, and these data points,
suitably weighted to reflect their imprecision, may yet provide some leverage in 
dynamical analyses.  As is shown in the separate panels of Figure 6, where the broken lines
represent the identity relationship, there is moderately good
systematic
agreement between the velocities compiled in Mould et al. (1990) and those determined separately
by Cohen and Ryzhov (1997) and in the present study. In the lower panel in particular, the formal
regression solution is \(v_{MOS} = 0.76\: (v_{M90}) + 280\: km\: sec^{-1};\) it is
shown as a solid line.   

\begin{figure*}[t]
\centering \leavevmode
\epsfysize=4.0truein
\epsfbox{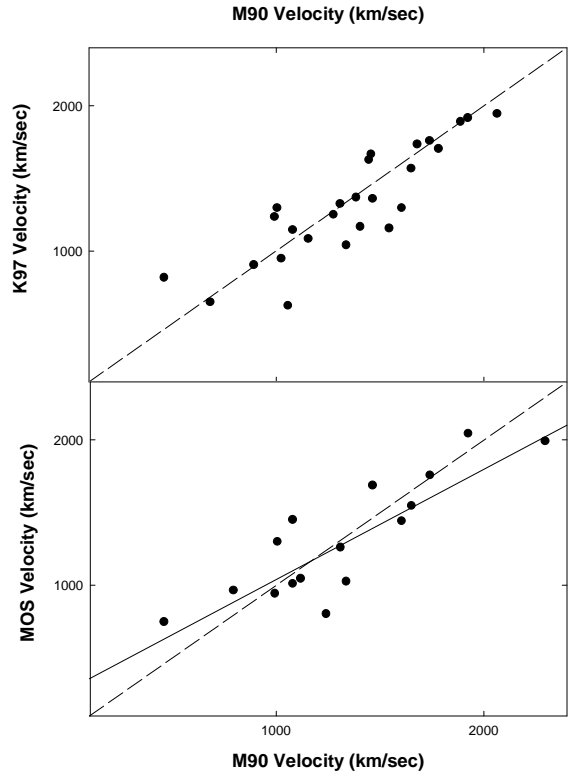}
\caption{A comparison of the velocities tabulated by Mould et al. (1990) with those of the present
study and those reported by Cohen and Ryzhov (1997) for objects in common.  In each panel, the broken line
is the identity relationship. In the lower panel, the  solid line represents the 
formal best-fit linear regression.
\label{fig6}}
\end{figure*}

\section{The Final Dataset} \label{Velocities}

In Figure 7, we plot 
the spatial distribution of
{\sl all objects for which velocities are now known in the M87 field,} clearly distiguished 
to represent the source of the information.  (In C\^ot\'e et al. (2001), we
will of course restrict the dynamical analysis to just that [large] fraction of the 
sample which are {\sl bona fide} globular clusters associated with M87.)  As can be seen,
the new MOS data are important
not just insofar as they augment the database by some $40\%$, but also---and just as
importantly---in that they extend the spatial coverage to larger galactocentric radii, a 
contribution which will be important in providing strong leverage in studies
of the dynamical state of the globular cluster system of M87 (C\^ot\'e et al. 2001).
Unfortunately, as the figure makes clear, there is a paucity of datapoints in the 
Northeast quadrant.  Nevertheless, the combination of photometric information,
sufficiently precise velocities, and {\em sheer numbers} of dynamical tracers will allow an
analytic treatment of unprecedented significance in this important subject area. 

\section{Conclusions}

We have presented a comprehensive dataset which incorporates all known velocity
tracers in the field of M87.  The new observations described here have augmented the 
results of earlier studies by $40\%$ in number (to a total of 352 objects)
and have been shown to be in excellent systematic agreement with the earlier
measurements.  We 
have extended the spatial
scale of the investigation to larger galactocentric radii than was formerly the case,
and have also been able
to provide precise CCD photometry in the Washington system for all but a few of
the objects.  Particular attention has been paid to the question of the 
unambiguous identification of all targets, the coordinates of which have
moreover been put onto an absolute astrometric scale.  In C\^ot\'e et al. (2001),
we present a dynamical study of the
GCS of M87 and describe the implications and inferences in
the context of galaxy formation and interaction history.

\acknowledgements

The research of DAH and GLHH is supported through grants from the Natural Sciences and 
Engineering Research Council of Canada.  DAH is pleased to thank the Directors of the
Dominion Astrophysical Observatory and the Anglo-Australian Observatory for their
hospitality and support during a research sabbatical.  PC gratefully acknowledges support
provided by the Sherman M. Fairchild Foundation during the course of this work.  DEM
acknowledges support from NASA through grant number HF-1097.01-97A awarded by the Space
Science Telescope Institute, which is operated by the Association of Universities for
Research in Astronomy, Inc., for NASA under contract NAS5-26555. DG acknowledges financial
support for this project received from CONICYT through Fundecyt grant 8000002, and from
the Universidad de Concepci\'on through research grant No. 99.011.025-1.0.

\begin{figure*}[b]
\vskip3.75truein
\caption{The spatial distribution of objects of known velocity in the M87 field; North is
up and East is to the left. The green filled circles represent new CFHT MOS measurements;
the blue squares represent repeat measurements of previous determinations; and the red
triangles  represent all other published determinations.
\label{fig7}}
\end{figure*}

\clearpage																					
\begin{deluxetable}{rrrccrrcccccc}
\tablenum{1}
\tablefontsize{\tiny}
\tablecolumns{13} 
\tablewidth{0pt}
\tablecaption{Positions, Photometry and Kinematics.}
\tablehead
{
   \colhead{ID}
 & \colhead{SX$^{"}$}
 & \colhead{SY$^{"}$}
 & \colhead{RA(2000)} 
 & \colhead{Dec(2000)} 
 & \colhead{$\Delta\alpha^{"}$}
 & \colhead{$\Delta\delta^{"}$} 
 & \colhead{T$_1$(mag)}
 & \colhead{C-T$_1$(mag)} 
 & \colhead{MOS (kms$^{-1}$)}
 & \colhead{K97 (kms$^{-1}$)}
 & \colhead{K00 (kms$^{-1}$)} 
 & \colhead{M90 (kms$^{-1}$)}
}
\startdata																					
9	&	430.2	&	 10.7	&	12		31		17.7	&	12		23		45	&	415	&	17	&	20.06	&	1.58	&	1324	$\pm$	68	&	\nodata	&	\nodata	&	\nodata	\\	
16	&	 89.7	&	 15.2	&	12		31		18.0	&	12		18		04	&	419	&	-324	&	21.55	&	1.45	&	\nodata		 	&	\nodata	&	1115	&	\nodata	\\	
28	&	140.7	&  	 22.7	&	12		31		17.4	&	12		18		55	&	410	&	-273	&	20.03	&	0.91	&	1253	$\pm$	88	&	\nodata	&	1157	&	\nodata	\\	
38	&	416.9	& 	 33.4	&	12		31		16.2	&	12		23		31	&	393	&	3	&	19.84	&	2.80	&	-96	$\pm$	128	&	not given	&	\nodata	&	\nodata	\\	
41	&	131.5	&	 38.7	&	12		31		16.3	&	12		18		45	&	395	&	-283	&	20.48	&	1.14	&	1760	$\pm$	111	&	\nodata	&	1814	&	\nodata	\\	
58	&	577.1	&	 52.4	&	12		31		14.7	&	12		26		11	&	370	&	163	&	21.16	&	0.87	&	\nodata		 	&	1923	&	\nodata	&	\nodata	\\	
59	&	439.8	&	 53.3	&	12		31		14.8	&	12		23		54	&	372	&	26	&	18.34	&	1.69	&	\nodata		 	&	-28	&	\nodata	&	\nodata	\\	
66	&	 62.9	&	 57.7	&	12		31		15.1	&	12		17		36	&	377	&	-352	&	20.45	&	1.13	&	2269	$\pm$	119	&	2260	&	2212	&	\nodata	\\	
77	&	311.9	&	 65.4	&	12		31		14.2	&	12		21		45	&	363	&	-103	&	19.78	&	1.14	&	1827	$\pm$	75	&	\nodata	&	1774	&	\nodata	\\	
87	&	 22.9	&	 73.1	&	12		31		14.2	&	12		16		56	&	363	&	-392	&	18.63	&	3.22	&	83	$\pm$	103	&	not given	&	\nodata	&	\nodata	\\	
91	&	425.6	&	 74.3	&	12		31		13.4	&	12		23		39	&	352	&	11	&	21.14	&	0.76	&	\nodata		 	&	179	&	\nodata	&	\nodata	\\	
94	&	244.1	& 	 77.6	&	12		31		13.5	&	12		20		37	&	353	&	-171	&	20.76	&	1.01	&	\nodata		 	&	\nodata	&	963	&	\nodata	\\	
101	&	453.1	&	 83.6	&	12		31		12.7	&	12		24		06	&	342	&	38	&	20.94	&	1.28	&	\nodata		 	&	1332	&	\nodata	&	\nodata	\\	
103	&	271.0	&	 83.8	&	12		31		13.0	&	12		21		04	&	346	&	-144	&	20.74	&	1.84	&	1204	$\pm$	182	&	\nodata	&	1092	&	\nodata	\\	
107	&	422.8	&  	 89.1	&	12		31		12.4	&	12		23		36	&	337	&	8	&	20.26	&	1.39	&	\nodata		 	&	1515	&	\nodata	&	\nodata	\\	
137	&	237.9	&	105.6	&	12		31		11.6	&	12		20		30	&	325	&	-178	&	19.64	&	1.10	&	\nodata		 	&	\nodata	&	1764	&	\nodata	\\	
141	&	383.2	&	109.1	&	12		31		11.1	&	12		22		56	&	318	&	-32	&	20.06	&	1.46	&	1207	$\pm$	91	&	1115	&	\nodata	&	\nodata	\\	
159	&	397.7	&	118.4	&	12		31		10.4	&	12		23		10	&	308	&	-18	&	21.28	&	1.54	&	1145	$\pm$	123	&	\nodata	&	\nodata	&	\nodata	\\	
161	&	173.1	&	119.3	&	12		31		10.8	&	12		19		25	&	313	&	-243	&	20.42	&	3.31	&	446	$\pm$	246	&	\nodata	&	\nodata	&	\nodata	\\	
176	&	 21.3	&	128.1	&	12		31		10.4	&	12		16		53	&	308	&	-395	&	20.31	&	1.43	&	2417	$\pm$	139	&	2210	&	2167	&	\nodata	\\	
177	& 	 48.8	&	129.9	&	12		31		10.2	&	12		17		20	&	305	&	-368	&	20.63	&	1.34	&	\nodata		 	&	1629	&	1540	&	1446	\\	
186	&	523.1	&	134.6	&	12		31		09.1	&	12		25		15	&	289	&	107	&	20.83	&	1.19	&	\nodata		 	&	1761	&	\nodata	&	\nodata	\\	
191	&	369.9	&	136.6	&	12		31		09.3	&	12		22		42	&	291	&	-46	&	20.87	&	1.06	&	\nodata		 	&	716	&	\nodata	&	\nodata	\\	
199	&	387.4	&	140.8	&	12		31		08.9	&	12		22		59	&	286	&	-29	&	19.71	&	3.52	&	0	$\pm$	120	&	\nodata	&	\nodata	&	\nodata	\\	
207	&	817.2	&	145.3	&	12		31		07.9	&	12		30		09	&	271	&	401	&	18.88	&	1.09	&	\nodata		 	&	\nodata	&	\nodata	&	1645	\\	
235	&	537.2	&	162.2	&	12		31		07.2	&	12		25		28	&	261	&	120	&	20.61	&	0.61	&	\nodata		 	&	31	&	\nodata	&	\nodata	\\	
248	&	464.0	&	166.7	&	12		31		07.0  &	12		24		15	&	258	&	47	&	19.89	&	1.32	&	\nodata		 	&	1016	&	\nodata	&	\nodata	\\	
252	&	592.2	&	169.2	&	12		31		06.7	&	12		26		23	&	253	&	175	&	21.23	&	0.74	&	\nodata		 	&	-41	&	\nodata	&	\nodata	\\	
270	&	 22.2	&	176.0	&	12		31		07.1	&	12		16		53	&	260	&	-395	&	18.25	&	1.87	&	112	$\pm$	47	&	\nodata	&	\nodata	&	\nodata	\\	
279	&	609.8	&	181.7	&	12		31		05.8	&	12		26		40	&	240	&	192	&	19.78	&	1.41	&	\nodata		 	&	820	&	\nodata	&	\nodata	\\	
280	&	214.4	&	182.8	&	12		31		06.3	&	12		20		05	&	248	&	-203	&	20.23	&	1.50	&	1298	$\pm$	66	&	1044	&	\nodata	&	\nodata	\\	
285	&	330.0	&	185.2	&	12		31		06.0	&	12		22		00	&	243	&	-88	&	19.75	&	1.30	&	1536	$\pm$	133	&	\nodata	&	\nodata	&	\nodata	\\	
286	&	192.2	&	185.3	&	12		31		06.2	&	12		19		42	&	246	&	-226	&	19.83	&	0.89	&	\nodata		 	&	12	&	\nodata	&	\nodata	\\	
290	&	234.8	&	187.5	&	12		31		06.0	&	12		20		25	&	243	&	-183	&	20.57	&	1.60	&	\nodata		 	&	1413	&	\nodata	&	\nodata	\\	
291	&	432.5	&	188.3	&	12		31		05.6	&	12		23		43	&	238	&	15	&	21.03	&	1.42	&	1781	$\pm$	115	&	\nodata	&	\nodata	&	\nodata	\\	
292	&	505.0	&	189.5	&	12		31		05.4	&	12		24		55	&	235	&	87	&	20.89	&	1.19	&	\nodata		 	&	847	&	\nodata	&	\nodata	\\	
300	&	 32.6	&	196.2	&	12		31		05.7	&	12		17		03	&	239	&	-385	&	20.66	&	1.05	&	1993	$\pm$	90	&	\nodata	&	1977	&	2296	\\	
307	&	130.8	&	198.3	&	12		31		05.4	&	12		18		41	&	235	&	-287	&	19.97	&	1.29	&	\nodata		 	&	1219	&	\nodata	&	\nodata	\\	
311	&	444.0	&	200.3	&	12		31		04.8	&	12		23		54	&	225	&	26	&	20.66	&	1.50	&	\nodata		 	&	781	&	\nodata	&	\nodata	\\	
313	&	147.9	&	200.9	&	12		31		05.2	&	12		18		58	&	232	&	-270	&	20.83	&	1.23	&	\nodata		 	&	1768	&	\nodata	&	\nodata	\\	
314	&	213.7	&	201.0	&	12		31		05.1	&	12		20		04	&	230	&	-204	&	19.46	&	1.68	&	1321	$\pm$	73	&	1236	&	1167	&	\nodata	\\	
321	&	615.1	&	206.0	&	12		31		04.1	&	12		26		45	&	216	&	197	&	19.95	&	1.04	&	-137	$\pm$	86	&	1376	&	\nodata	&	\nodata	\\	
323	&	411.2	&	206.6	&	12		31		04.4	&	12		23		21	&	220	&	-7	&	20.07	&	1.14	&	\nodata		 	&	1124	&	\nodata	&	\nodata	\\	
324	&	 98.7	&	206.6	&	12		31		04.9	&	12		18		08	&	227	&	-320	&	20.45	&	1.37	&	\nodata		 	&	359	&	\nodata	&	\nodata	\\	
330	&	620.2	&	211.4	&	12		31		03.7	&	12		26		50	&	210	&	202	&	20.82	&	1.52	&	\nodata		 	&	736	&	\nodata	&	\nodata	\\	
339	&	424.5	&	217.1	&	12		31		03.7	&	12		23		34	&	209	&	6	&	21.14	&	1.40	&	-574	$\pm$	120	&	\nodata	&	\nodata	&	\nodata	\\	
348	&	396.5	&	222.2	&	12		31		03.4	&	12		23		06	&	205	&	-22	&	19.09	&	1.30	&	748	$\pm$	102	&	817	&	\nodata	&	460	\\	
350	&	640.8	&	222.9	&	12		31		02.9	&	12		27		10	&	198	&	222	&	20.80	&	1.15	&	\nodata		 	&	1244	&	\nodata	&	\nodata	\\	
357	&	659.0	&	224.9	&	12		31		02.8	&	12		27		29	&	196	&	241	&	21.52	&	0.44	&	\nodata		 	&	-86	&	\nodata	&	\nodata	\\	
361	&	437.8	&	228.9	&	12		31		02.8	&	12		23		47	&	197	&	19	&	20.90	&	1.20	&	1529	$\pm$	138	&	\nodata	&	\nodata	&	\nodata	\\	
376	&	680.0	&	233.6	&	12		31		02.1	&	12		27		49	&	186	&	261	&	20.08	&	1.05	&	\nodata		 	&	1182	&	\nodata	&	\nodata	\\	
378	&	417.5	&	233.7	&	12		31		02.5	&	12		23		27	&	193	&	-1	&	20.29	&	1.29	&	\nodata		 	&	1978	&	\nodata	&	\nodata	\\	
379	&	 60.0	&	234.8	&	12		31		03.0	&	12		17		29	&	200	&	-359	&	20.11	&	1.14	&	\nodata		 	&	\nodata	&	2244	&	\nodata	\\	
395	&	404.3	&	243.0	&	12		31		01.9	&	12		23		13	&	184	&	-15	&	21.10	&	1.11	&	\nodata		 	&	1890	&	\nodata	&	\nodata	\\	
409	&	412.9	&	250.9	&	12		31		01.4	&	12		23		22	&	175	&	-6	&	21.37	&	1.44	&	2197	$\pm$	103	&	\nodata	&	\nodata	&	\nodata	\\	
415	&	 63.1	&	254.1	&	12		31		01.7	&	12		17		32	&	181	&	-356	&	20.61	&	1.51	&	\nodata		 	&	\nodata	&	1817	&	\nodata	\\	
417	&	176.2	&	255.4	&	12		31		01.5	&	12		19		25	&	177	&	-243	&	18.81	&	1.59	&	\nodata		 	&	1910	&	\nodata	&	\nodata	\\	
418	&	268.0	&	256.3	&	12		31		01.2	&	12		20		57	&	174	&	-151	&	20.48	&	1.22	&	\nodata		 	&	1866	&	\nodata	&	\nodata	\\	
420	&	725.4	&	256.8	&	12		31		00.5	&	12		28		34	&	162	&	306	&	20.15	&	0.68	&	\nodata		 	&	-153	&	\nodata	&	\nodata	\\	
421	&	498.0	&	257.0	&	12		31		00.8	&	12		24		47	&	167	&	79	&	19.74	&	1.58	&	1511	$\pm$	74	&	1724	&	\nodata	&	\nodata	\\	
423	&	226.4	&	258.7	&	12		31		01.1	&	12		20		15	&	172	&	-193	&	20.87	&	1.60	&	\nodata		 	&	1081	&	\nodata	&	\nodata	\\	
442	&	729.2	&	266.7	&	12		30		59.8	&	12		28		38	&	152	&	310	&	20.00	&	1.32	&	\nodata		 	&	1424	&	\nodata	&	\nodata	\\	
453	&	426.3	&	270.8	&	12		31		00.0	&	12		23		35	&	155	&	7	&	20.56	&	1.38	&	\nodata		 	&	1980	&	\nodata	&	\nodata	\\	
468	&	614.9	&	275.3	&	12		30		59.4	&	12		26		43	&	146	&	195	&	20.03	&	1.88	&	802	$\pm$	93	&	\nodata	&	\nodata	&	1240	\\	
477	&	 93.2	&	280.3	&	12		30		59.9	&	12		18		01	&	154	&	-327	&	19.39	&	1.12	&	\nodata		 	&	\nodata	&	1584	&	1633	\\	
490	&	295.3	&	283.6	&	12		30		59.3	&	12		21		23	&	146	&	-125	&	19.53	&	1.98	&	1549	$\pm$	56	&	1570	&	\nodata	&	1651	\\	
491	&	284.4	&	283.8	&	12		30		59.3	&	12		21		12	&	146	&	-136	&	20.54	&	1.23	&	\nodata		 	&	1069	&	\nodata	&	\nodata	\\	
492	&	462.2	&	284.0	&	12		30		59.0	&	12		24		10	&	141	&	42	&	20.84	&	1.49	&	\nodata		 	&	1498	&	\nodata	&	\nodata	\\	
501	&	596.4	&	287.7	&	12		30		58.6	&	12		26		24	&	134	&	176	&	19.69	&	1.32	&	\nodata		 	&	\nodata	&	\nodata	&	1940	\\	
508	&	591.0	&	291.1	&	12		30		58.3	&	12		26		19	&	131	&	171	&	19.48	&	3.03	&	5817	$\pm$	120	&	\nodata	&	\nodata	&	\nodata	\\	
514	&	 35.0	&	293.7	&	12		30		59.1	&	12		17		02	&	142	&	-386	&	21.03	&	1.10	&	\nodata		 	&	\nodata	&	1617	&	\nodata	\\	
518	&	758.0	&	296.7	&	12		30		57.7	&	12		29		06	&	121	&	338	&	20.79	&	0.57	&	\nodata		 	&	42	&	\nodata	&	\nodata	\\	
519	&	255.3	&	296.9	&	12		30		58.5	&	12		20		43	&	133	&	-165	&	20.86	&	1.58	&	\nodata		 	&	1315	&	\nodata	&	\nodata	\\	
537	&	232.9	&	303.7	&	12		30		58.1	&	12		20		20	&	127	&	-188	&	20.53	&	1.21	&	\nodata		 	&	1467	&	\nodata	&	\nodata	\\	
547	&	556.6	&	306.2	&	12		30		57.4	&	12		25		44	&	117	&	136	&	17.88	&	1.95	&	\nodata		 	&	\nodata	&	\nodata	&	758	\\	
571	&	412.0	&	312.6	&	12		30		57.2	&	12		23		19	&	114	&	-9	&	20.51	&	1.50	&	\nodata		 	&	1757	&	\nodata	&	\nodata	\\	
572	&	292.4	&	313.2	&	12		30		57.3	&	12		21		20	&	116	&	-128	&	19.96	&	2.71	&	125	$\pm$	85	&	\nodata	&	\nodata	&	\nodata	\\	
579	&	280.2	&	315.5	&	12		30		57.2	&	12		21		07	&	114	&	-141	&	20.50	&	1.14	&	\nodata		 	&	999	&	\nodata	&	\nodata	\\	
581	&	330.3	&	316.8	&	12		30		57.0	&	12		21		57	&	112	&	-91	&	20.43	&	1.47	&	\nodata		 	&	1507	&	\nodata	&	\nodata	\\	
582	&	603.4	&	317.4	&	12		30		56.5	&	12		26		31	&	104	&	183	&	19.87	&	1.24	&	\nodata		 	&	\nodata	&	\nodata	&	1530	\\	
588	&	280.8	&	321.1	&	12		30		56.8	&	12		21		08	&	108	&	-140	&	19.84	&	1.33	&	1485	$\pm$	99	&	1637	&	\nodata	&	\nodata	\\	
602	&	424.0	&	327.2	&	12		30		56.1	&	12		23		31	&	99	&	3	&	19.67	&	1.53	&	688	$\pm$	88	&	555	&	\nodata	&	\nodata	\\	
603	&	 40.9	&	327.2	&	12		30		56.8	&	12		17		08	&	108	&	-380	&	21.01	&	1.06	&	\nodata		 	&	\nodata	&	1741	&	\nodata	\\	
611	&	627.1	&	328.9	&	12		30		55.7	&	12		26		54	&	92	&	206	&	20.66	&	1.18	&	\nodata		 	&	1319	&	\nodata	&	\nodata	\\	
612	&	444.2	&	328.9	&	12		30		56.0	&	12		23		51	&	97	&	23	&	19.94	&	1.55	&	1444	$\pm$	136	&	\nodata	&	\nodata	&	\nodata	\\	
614	&	749.1	&	329.2	&	12		30		55.5	&	12		28		56	&	89	&	328	&	20.77	&	1.29	&	\nodata		 	&	1891	&	\nodata	&	\nodata	\\	
645	&	369.3	&	342.5	&	12		30		55.2	&	12		22		36	&	85	&	-52	&	19.37	&	1.69	&	1758	$\pm$	123	&	1760	&	\nodata	&	1740	\\	
647	&	524.5	&	342.7	&	12		30		54.9	&	12		25		11	&	81	&	103	&	19.90	&	1.38	&	923	$\pm$	121	&	972	&	\nodata	&	\nodata	\\	
649	&	412.1	&	343.3	&	12		30		55.1	&	12		23		19	&	83	&	-9	&	19.95	&	1.33	&	\nodata		 	&	1373	&	\nodata	&	\nodata	\\	
651	&	267.9	&	344.1	&	12		30		55.2	&	12		20		54	&	86	&	-154	&	20.70	&	1.74	&	\nodata		 	&	2110	&	\nodata	&	\nodata	\\	
663	&	254.3	&	348.6	&	12		30		55.0	&	12		20		41	&	82	&	-167	&	18.63	&	1.16	&	150	$\pm$	90	&	\nodata	&	\nodata	&	\nodata	\\	
664	&	341.1	&	348.6	&	12		30		54.8	&	12		22		07	&	79	&	-81	&	21.04	&	1.33	&	\nodata		 	&	1563	&	\nodata	&	\nodata	\\	
668	&	577.6	&	349.7	&	12		30		54.4	&	12		26		04	&	73	&	156	&	20.16	&	1.42	&	\nodata		 	&	\nodata	&	\nodata	&	1510	\\	
672	&	312.3	&	350.9	&	12		30		54.7	&	12		21		39	&	78	&	-109	&	20.15	&	1.24	&	\nodata		 	&	702	&	\nodata	&	\nodata	\\	
678	&	108.9	&	353.5	&	12		30		54.9	&	12		18		15	&	80	&	-313	&	21.12	&	0.65	&	\nodata		 	&	-94	&	\nodata	&	\nodata	\\	
679	&	780.4	&	353.5	&	12		30		53.8	&	12		29		27	&	64	&	359	&	20.61	&	1.49	&	\nodata		 	&	1218	&	\nodata	&	\nodata	\\	
680	&	494.3	&	353.6	&	12		30		54.2	&	12		24		41	&	71	&	73	&	20.55	&	1.21	&	\nodata		 	&	1793	&	\nodata	&	\nodata	\\	
682	&	236.6	&	353.7	&	12		30		54.6	&	12		20		23	&	77	&	-185	&	20.69	&	1.10	&	\nodata		 	&	1300	&	\nodata	&	\nodata	\\	
684	&	608.7	&	353.9	&	12		30		54.0	&	12		26		35	&	68	&	187	&	20.11	&	1.20	&	\nodata		 	&	\nodata	&	\nodata	&	1430	\\	
686	&	711.8	&	354.7	&	12		30		53.8	&	12		28		18	&	64	&	290	&	19.84	&	1.17	&	\nodata		 	&	784	&	\nodata	&	\nodata	\\	
\enddata
\end{deluxetable}

\clearpage
\begin{deluxetable}{rrrccrrcccccc}
\tablenum{1}
\tablefontsize{\tiny}
\tablecolumns{13} 
\tablewidth{0pt}
\tablecaption{Positions, Photometry and Kinematics.}
\tablehead
{
   \colhead{ID}
 & \colhead{SX$^{"}$}
 & \colhead{SY$^{"}$}
 & \colhead{RA(2000)} 
 & \colhead{Dec(2000)} 
 & \colhead{$\Delta\alpha^{"}$}
 & \colhead{$\Delta\delta^{"}$} 
 & \colhead{T$_1$(mag)}
 & \colhead{C-T$_1$(mag)} 
 & \colhead{MOS (kms$^{-1}$)}
 & \colhead{K97 (kms$^{-1}$)}
 & \colhead{K00 (kms$^{-1}$)} 
 & \colhead{M90 (kms$^{-1}$)}
}
\startdata
695	&	376.3	&	357.1	&	12		30		54.2	&	12		22		42	&	70	&	-46	&	20.62	&	1.30	&	\nodata		 	&	1650	&	\nodata	&	\nodata	\\	
697	&	388.8	&	358.1	&	12		30		54.1	&	12		22		55	&	69	&	-33	&	19.89	&	1.78	&	\nodata		 	&	1203	&	\nodata	&	\nodata	\\	
708	&	589.3	&	362.4	&	12		30		53.5	&	12		26		15	&	60	&	167	&	20.62	&	1.70	&	\nodata		 	&	\nodata	&	\nodata	&	1130	\\	
714	&	366.6	&	363.5	&	12		30		53.8	&	12		22		33	&	64	&	-55	&	20.31	&	1.25	&	1254	$\pm$	122	&	1315	&	\nodata	&	\nodata	\\	
720	&	561.2	&	365.4	&	12		30		53.3	&	12		25		47	&	57	&	139	&	19.14	&	2.68	&	-25	$\pm$	120	&	\nodata	&	\nodata	&	\nodata	\\	
723	&	235.9	&	367.1	&	12		30		53.7	&	12		20		22	&	63	&	-186	&	21.06	&	1.20	&	\nodata		 	&	1365	&	\nodata	&	\nodata	\\	
731	&	046.8	&	369.5	&	12		30		53.9	&	12		17		12	&	66	&	-376	&	20.35	&	1.33	&	\nodata		 	&	\nodata	&	960	&	\nodata	\\	
741	&	266.6	&	374.0	&	12		30		53.2	&	12		20		52	&	56	&	-156	&	20.23	&	1.75	&	1269	$\pm$	135	&	1212	&	\nodata	&	\nodata	\\	
746	&	476.9	&	375.7	&	12		30		52.8	&	12		24		23	&	49	&	55	&	19.27	&	1.59	&	\nodata		 	&	1266	&	\nodata	&	\nodata	\\	
750	&	361.3	&	376.7	&	12		30		52.9	&	12		22		27	&	51	&	-61	&	20.11	&	1.35	&	\nodata		 	&	1406	&	\nodata	&	\nodata	\\	
762	&	683.2	&	382.6	&	12		30		51.9	&	12		27		49	&	37	&	261	&	21.47	&	1.94	&	1177	$\pm$	115	&	\nodata	&	\nodata	&	\nodata	\\	
766	&	819.7	&	384.0	&	12		30		51.6	&	12		30		06	&	33	&	398	&	20.41	&	1.18	&	1107	$\pm$	83	&	\nodata	&	\nodata	&	\nodata	\\	
770	&	504.8	&	385.1	&	12		30		52.1	&	12		24		50	&	39	&	82	&	20.35	&	1.41	&	\nodata		 	&	1481	&	\nodata	&	\nodata	\\	
782	&	204.5	&	390.2	&	12		30		52.2	&	12		19		50	&	41	&	-218	&	18.96	&	2.98	&	-26	$\pm$	126	&	not given	&	\nodata	&	\nodata	\\	
784	&	621.9	&	391.3	&	12		30		51.4	&	12		26		47	&	30	&	199	&	19.35	&	1.46	&	\nodata		 	&	1891	&	\nodata	&	1888	\\	
787	&	193.1	&	392.8	&	12		30		52.0	&	12		19		38	&	39	&	-230	&	20.64	&	1.34	&	\nodata		 	&	1097	&	\nodata	&	\nodata	\\	
789	&	524.5	&	393.6	&	12		30		51.5	&	12		25		10	&	30	&	102	&	20.47	&	1.22	&	966	$\pm$	71	&	\nodata	&	\nodata	&	795	\\	
796	&	268.3	&	395.5	&	12		30		51.7	&	12		20		53	&	34	&	-155	&	20.10	&	1.28	&	\nodata		 	&	1130	&	\nodata	&	\nodata	\\	
798	&	242.2	&	395.9	&	12		30		51.8	&	12		20		27	&	34	&	-181	&	20.31	&	1.26	&	859	$\pm$	173	&	950	&	\nodata	&	\nodata	\\	
801	&	766.8	&	396.4	&	12		30		50.9	&	12		29		12	&	21	&	344	&	18.44	&	3.65	&	\nodata		 	&	-30	&	\nodata	&	\nodata	\\	
804	&	586.2	&	397.8	&	12		30		51.1	&	12		26		12	&	24	&	164	&	18.84	&	1.55	&	1012	$\pm$	71	&	\nodata	&	\nodata	&	1079	\\	
807	&	527.1	&	399.1	&	12		30		51.1	&	12		25		12	&	24	&	104	&	20.46	&	1.42	&	745	$\pm$	194	&	\nodata	&	\nodata	&	\nodata	\\	
809	&	182.7	&	400.1	&	12		30		51.6	&	12		19		28	&	32	&	-240	&	20.04	&	1.35	&	\nodata		 	&	612	&	\nodata	&	\nodata	\\	
811	&	321.0	&	401.0	&	12		30		51.3	&	12		21		46	&	27	&	-102	&	20.22	&	1.11	&	\nodata		 	&	1436	&	\nodata	&	\nodata	\\	
814	&	721.2	&	401.6	&	12		30		50.6	&	12		28		27	&	17	&	299	&	21.33	&	1.41	&	\nodata		 	&	1331	&	\nodata	&	\nodata	\\	
824	&	341.5	&	405.3	&	12		30		50.9	&	12		22		06	&	23	&	-82	&	20.33	&	1.40	&	\nodata		 	&	1212	&	\nodata	&	\nodata	\\	
825	&	294.6	&	405.4	&	12		30		51.0	&	12		21		19	&	24	&	-129	&	21.00	&	1.18	&	\nodata		 	&	1109	&	\nodata	&	\nodata	\\	
829	&	543.7	&	408.5	&	12		30		50.4	&	12		25		29	&	15	&	121	&	20.47	&	1.24	&	\nodata		 	&	\nodata	&	\nodata	&	835	\\	
831	&	660.7	&	409.9	&	12		30		50.1	&	12		27		26	&	10	&	238	&	20.07	&	1.21	&	1451	$\pm$	105	&	1148	&	\nodata	&	1080	\\	
838	&	571.6	&	412.0	&	12		30		50.1	&	12		25		57	&	11	&	149	&	20.11	&	1.31	&	\nodata		 	&	1086	&	\nodata	&	1155	\\	
849	&	166.3	&	416.4	&	12		30		50.5	&	12		19		11	&	16	&	-257	&	20.46	&	1.62	&	1771	$\pm$	110	&	1650	&	\nodata	&	\nodata	\\	
857	&	482.7	&	420.5	&	12		30		49.7	&	12		24		27	&	4	&	59	&	20.51	&	1.38	&	\nodata		 	&	759	&	\nodata	&	\nodata	\\	
859	&	673.1	&	422.1	&	12		30		49.3	&	12		27		38	&	-2	&	250	&	20.98	&	1.30	&	1064	$\pm$	91	&	\nodata	&	\nodata	&	\nodata	\\	
871	&	227.2	&	428.3	&	12		30		49.6	&	12		20		11	&	2	&	-197	&	20.79	&	1.61	&	\nodata		 	&	1073	&	\nodata	&	\nodata	\\	
878	&	 29.0	&	431.3	&	12		30		49.7	&	12		16		53	&	4	&	-395	&	20.66	&	1.33	&	2512	$\pm$	98	&	\nodata	&	\nodata	&	\nodata	\\	
879	&	346.3	&	431.5	&	12		30		49.2	&	12		22		11	&	-4	&	-77	&	20.96	&	1.47	&	1806	$\pm$	103	&	\nodata	&	\nodata	&	\nodata	\\	
887	&	331.4	&	435.1	&	12		30		48.9	&	12		21		56	&	-7	&	-92	&	20.37	&	1.47	&	\nodata		 	&	1778	&	\nodata	&	\nodata	\\	
902	&	494.3	&	439.9	&	12		30		48.3	&	12		24		38	&	-16	&	70	&	20.68	&	2.08	&	\nodata		 	&	1587	&	\nodata	&	\nodata	\\	
904	&	726.8	&	440.4	&	12		30		47.9	&	12		28		31	&	-22	&	303	&	20.58	&	1.48	&	\nodata		 	&	914	&	\nodata	&	\nodata	\\	
910	&	590.9	&	444.8	&	12		30		47.8	&	12		26		15	&	-23	&	167	&	20.13	&	1.15	&	\nodata		 	&	1007	&	\nodata	&	\nodata	\\	
917	&	 89.7	&	446.2	&	12		30		48.6	&	12		17		53	&	-12	&	-335	&	20.84	&	1.44	&	\nodata		 	&	882	&	\nodata	&	\nodata	\\	
922	&	764.2	&	447.1	&	12		30		47.4	&	12		29		08	&	-29	&	340	&	21.09	&	1.36	&	\nodata		 	&	1760	&	\nodata	&	\nodata	\\	
923	&	208.5	&	447.3	&	12		30		48.3	&	12		19		52	&	-16	&	-216	&	21.52	&	1.07	&	1943	$\pm$	134	&	\nodata	&	\nodata	&	\nodata	\\	
928	&	485.8	&	449.3	&	12		30		47.7	&	12		24		30	&	-25	&	62	&	19.02	&	1.31	&	1261	$\pm$	89	&	1327	&	\nodata	&	1309	\\	
937	&	659.3	&	452.1	&	12		30		47.2	&	12		27		23	&	-32	&	235	&	20.59	&	1.22	&	\nodata		 	&	930	&	\nodata	&	\nodata	\\	
941	&	674.7	&	453.1	&	12		30		47.1	&	12		27		39	&	-33	&	251	&	19.99	&	1.23	&	\nodata		 	&	1140	&	\nodata	&	\nodata	\\	
946	&	524.2	&	455.1	&	12		30		47.3	&	12		25		08	&	-31	&	100	&	20.48	&	1.55	&	\nodata		 	&	1131	&	\nodata	&	\nodata	\\	
947	&	 66.4	&	455.9	&	12		30		47.9	&	12		17		30	&	-21	&	-358	&	21.05	&	1.25	&	\nodata		 	&	1519	&	\nodata	&	\nodata	\\	
949	&	699.1	&	457.3	&	12		30		46.8	&	12		28		03	&	-38	&	275	&	21.20	&	1.63	&	472	$\pm$	90	&	\nodata	&	\nodata	&	\nodata	\\	
952	&	 14.6	&	458.3	&	12		30		47.9	&	12		16		38	&	-23	&	-410	&	20.63	&	1.18	&	\nodata		 	&	1454	&	\nodata	&	\nodata	\\	
963	&	831.4	&	464.1	&	12		30		46.1	&	12		30		15	&	-48	&	407	&	20.11	&	1.28	&	633	$\pm$	107	&	\nodata	&	\nodata	&	\nodata	\\	
965	&	179.7	&	465.2	&	12		30		47.1	&	12		19		23	&	-33	&	-245	&	20.07	&	1.54	&	\nodata		 	&	1362	&	\nodata	&	1465	\\	
968	&	287.0	&	466.2	&	12		30		46.9	&	12		21		10	&	-37	&	-138	&	21.02	&	1.32	&	\nodata		 	&	1091	&	\nodata	&	\nodata	\\	
970	&	538.0	&	466.9	&	12		30		46.4	&	12		25		22	&	-44	&	114	&	20.11	&	1.49	&	\nodata		 	&	991	&	\nodata	&	\nodata	\\	
973	&	255.3	&	468.0	&	12		30		46.8	&	12		20		39	&	-38	&	-169	&	20.52	&	0.81	&	\nodata		 	&	172	&	\nodata	&	\nodata	\\	
991	&	234.6	&	473.1	&	12		30		46.5	&	12		20		18	&	-43	&	-190	&	20.20	&	1.26	&	\nodata		 	&	950	&	\nodata	&	1025	\\	
992	&	681.2	&	473.6	&	12		30		45.7	&	12		27		45	&	-54	&	257	&	20.75	&	1.29	&	732	$\pm$	128	&	727	&	\nodata	&	\nodata	\\	
999	&	518.5	&	474.7	&	12		30		45.9	&	12		25		02	&	-51	&	94	&	19.56	&	1.29	&	\nodata		 	&	\nodata	&	\nodata	&	1515	\\	
1007	&	 59.7	&	478.8	&	12		30		46.4	&	12		17		23	&	-44	&	-365	&	19.53	&	1.68	&	1443	$\pm$	89	&	1298	&	\nodata	&	1604	\\	
1010	&	302.5	&	479.6	&	12		30		45.9	&	12		21		26	&	-51	&	-122	&	20.08	&	1.79	&	\nodata		 	&	1507	&	\nodata	&	\nodata	\\	
1015	&	535.5	&	480.8	&	12		30		45.5	&	12		25		19	&	-57	&	111	&	20.19	&	1.30	&	\nodata		 	&	1165	&	\nodata	&	\nodata	\\	
1016	&	311.4	&	481.3	&	12		30		45.8	&	12		21		34	&	-53	&	-114	&	19.92	&	2.11	&	1766	$\pm$	111	&	1375	&	\nodata	&	\nodata	\\	
1019	&	697.7	&	482.5	&	12		30		45.1	&	12		28		01	&	-63	&	273	&	19.33	&	1.32	&	\nodata		 	&	650	&	\nodata	&	683	\\	
1023	&	549.9	&	483.2	&	12		30		45.3	&	12		25		33	&	-60	&	125	&	20.57	&	1.43	&	\nodata		 	&	1164	&	\nodata	&	\nodata	\\	
1032	&	482.0	&	485.6	&	12		30		45.2	&	12		24		25	&	-61	&	57	&	20.23	&	1.45	&	1735	$\pm$	131	&	1684	&	\nodata	&	\nodata	\\	
1034	&	171.1	&	487.0	&	12		30		45.6	&	12		19		14	&	-55	&	-254	&	20.36	&	1.48	&	1301	$\pm$	113	&	1299	&	\nodata	&	1005	\\	
1044	&	250.7	&	489.7	&	12		30		45.3	&	12		20		33	&	-60	&	-175	&	20.03	&	1.29	&	\nodata		 	&	1947	&	\nodata	&	2065	\\	
1049	&	261.7	&	490.9	&	12		30		45.2	&	12		20		44	&	-61	&	-164	&	20.59	&	1.63	&	\nodata		 	&	1569	&	\nodata	&	\nodata	\\	
1055	&	674.1	&	492.8	&	12		30		44.4	&	12		27		37	&	-73	&	249	&	20.45	&	1.31	&	\nodata		 	&	1543	&	\nodata	&	\nodata	\\	
1060	&	514.2	&	493.6	&	12		30		44.6	&	12		24		57	&	-70	&	89	&	20.73	&	1.39	&	\nodata		 	&	1622	&	\nodata	&	\nodata	\\	
1064	&	507.9	&	494.6	&	12		30		44.6	&	12		24		51	&	-71	&	83	&	20.72	&	1.30	&	\nodata		 	&	1405	&	\nodata	&	\nodata	\\	
1070	&	800.6	&	495.9	&	12		30		44.0	&	12		29		44	&	-79	&	376	&	20.63	&	1.51	&	1370	$\pm$	90	&	1437	&	\nodata	&	\nodata	\\	
1074	&	 17.2	&	496.8	&	12		30		45.2	&	12		16		40	&	-61	&	-408	&	20.46	&	1.87	&	2186	$\pm$	137	&	\nodata	&	\nodata	&	\nodata	\\	
1091	&	444.6	&	506.7	&	12		30		43.9	&	12		23		47	&	-81	&	19	&	20.25	&	1.39	&	1027	$\pm$	94	&	1015	&	\nodata	&	\nodata	\\	
1093	&	700.3	&	507.4	&	12		30		43.4	&	12		28		03	&	-88	&	275	&	19.23	&	1.47	&	\nodata		 	&	905	&	\nodata	&	893	\\	
1101	&	486.5	&	508.2	&	12		30		43.7	&	12		24		29	&	-84	&	61	&	20.63	&	1.49	&	\nodata		 	&	1494	&	\nodata	&	\nodata	\\	
1103	&	528.3	&	509.2	&	12		30		43.6	&	12		25		11	&	-86	&	103	&	20.62	&	1.25	&	\nodata		 	&	-31	&	\nodata	&	\nodata	\\	
1108	&	200.6	&	510.7	&	12		30		44.0	&	12		19		43	&	-79	&	-225	&	20.17	&	1.45	&	1860	$\pm$	95	&	1930	&	\nodata	&	\nodata	\\	
1110	&	237.4	&	511.2	&	12		30		43.9	&	12		20		20	&	-81	&	-188	&	20.87	&	1.22	&	\nodata		 	&	1091	&	\nodata	&	\nodata	\\	
1112	&	444.3	&	512.0	&	12		30		43.5	&	12		23		47	&	-86	&	19	&	20.24	&	2.10	&	995	$\pm$	126	&	\nodata	&	\nodata	&	\nodata	\\	
1113	&	468.9	&	512.2	&	12		30		43.4	&	12		24		11	&	-87	&	43	&	20.46	&	1.19	&	\nodata		 	&	1544	&	\nodata	&	\nodata	\\	
1116	&	499.3	&	513.3	&	12		30		43.3	&	12		24		42	&	-89	&	74	&	20.33	&	1.52	&	\nodata		 	&	1084	&	\nodata	&	\nodata	\\	
1117	&	811.2	&	513.3	&	12		30		42.8	&	12		29		54	&	-96	&	386	&	20.53	&	1.65	&	1196	$\pm$	84	&	\nodata	&	\nodata	&	\nodata	\\	
1119	&	733.0	&	513.9	&	12		30		42.9	&	12		28		36	&	-95	&	308	&	20.68	&	1.48	&	\nodata		 	&	1510	&	\nodata	&	\nodata	\\	
1120	&	609.2	&	514.1	&	12		30		43.1	&	12		26		32	&	-93	&	184	&	20.80	&	1.25	&	1397	$\pm$	74	&	1367	&	\nodata	&	\nodata	\\	
1124	&	633.5	&	515.0	&	12		30		43.0	&	12		26		56	&	-94	&	208	&	20.14	&	1.49	&	1269	$\pm$	135	&	\nodata	&	\nodata	&	\nodata	\\	
1144	&	763.0	&	520.5	&	12		30		42.4	&	12		29		05	&	-103	&	337	&	21.00	&	1.10	&	\nodata		 	&	808	&	\nodata	&	\nodata	\\	
1155	&	561.6	&	523.3	&	12		30		42.5	&	12		25		44	&	-101	&	136	&	19.88	&	1.75	&	\nodata		 	&	1370	&	\nodata	&	1385	\\	
1157	&	422.0	&	523.8	&	12		30		42.7	&	12		23		24	&	-98	&	-4	&	19.67	&	1.37	&	1786	$\pm$	188	&	1731	&	\nodata	&	\nodata	\\	
1158	&	403.9	&	523.9	&	12		30		42.7	&	12		23		06	&	-97	&	-22	&	20.33	&	1.27	&	\nodata		 	&	1048	&	\nodata	&	\nodata	\\	
1165	&	597.9	&	525.8	&	12		30		42.3	&	12		26		20	&	-104	&	172	&	20.14	&	1.37	&	\nodata		 	&	1473	&	\nodata	&	\nodata	\\	
1167	&	723.1	&	526.1	&	12		30		42.1	&	12		28		25	&	-107	&	297	&	20.87	&	1.12	&	\nodata		 	&	1476	&	\nodata	&	\nodata	\\	
1173	&	361.9	&	529.2	&	12		30		42.5	&	12		22		24	&	-102	&	-64	&	20.30	&	1.85	&	1657	$\pm$	102	&	\nodata	&	\nodata	&	\nodata	\\	
1181	&	756.0	&	531.5	&	12		30		41.7	&	12		28		58	&	-113	&	330	&	20.64	&	1.45	&	\nodata		 	&	639	&	\nodata	&	\nodata	\\	
1200	&	471.2	&	536.6	&	12		30		41.8	&	12		24		13	&	-112	&	45	&	19.78	&	1.59	&	\nodata		 	&	878	&	\nodata	&	\nodata	\\	
1201	&	439.2	&	536.6	&	12		30		41.8	&	12		23		41	&	-111	&	13	&	20.48	&	1.23	&	\nodata		 	&	1178	&	\nodata	&	\nodata	\\	
1205	&	828.4	&	537.3	&	12		30		41.1	&	12		30		10	&	-121	&	402	&	20.04	&	1.27	&	1130	$\pm$	92	&	1011	&	\nodata	&	\nodata	\\	
1208	&	821.5	&	539.1	&	12		30		41.0	&	12		30		03	&	-123	&	395	&	20.88	&	1.96	&	\nodata		 	&	1605	&	\nodata	&	\nodata	\\	
1216	&	165.3	&	540.9	&	12		30		42.0	&	12		19		07	&	-109	&	-261	&	18.58	&	0.96	&	\nodata		 	&	101	&	\nodata	&	\nodata	\\	
1217	&	372.4	&	541.4	&	12		30		41.6	&	12		22		34	&	-114	&	-54	&	20.65	&	1.69	&	\nodata		 	&	1125	&	\nodata	&	\nodata	\\	
\enddata
\end{deluxetable}

\clearpage
\begin{deluxetable}{rrrccrrcccccc}
\tablenum{1}
\tablefontsize{\tiny}
\tablecolumns{13} 
\tablewidth{0pt}
\tablecaption{Positions, Photometry and Kinematics.}
\tablehead
{
   \colhead{ID}
 & \colhead{SX$^{"}$}
 & \colhead{SY$^{"}$}
 & \colhead{RA(2000)} 
 & \colhead{Dec(2000)} 
 & \colhead{$\Delta\alpha^{"}$}
 & \colhead{$\Delta\delta^{"}$} 
 & \colhead{T$_1$(mag)}
 & \colhead{C-T$_1$(mag)} 
 & \colhead{MOS (kms$^{-1}$)}
 & \colhead{K97 (kms$^{-1}$)}
 & \colhead{K00 (kms$^{-1}$)} 
 & \colhead{M90 (kms$^{-1}$)}
}
\startdata
1219	&	557.1	&	543.3	&	12		30		41.2	&	12		25		39	&	-120	&	131	&	20.45	&	1.55	&	\nodata		 	&	1244	&	\nodata	&	\nodata	\\	
1220	&	498.7	&	544.4	&	12		30		41.2	&	12		24		40	&	-120	&	72	&	20.27	&	1.35	&	970	$\pm$	86	&	855	&	\nodata	&	\nodata	\\	
1224	&	630.4	&	545.8	&	12		30		40.9	&	12		26		52	&	-125	&	204	&	21.06	&	1.48	&	1488	$\pm$	74	&	\nodata	&	\nodata	&	\nodata	\\	
1238	&	433.7	&	548.8	&	12		30		41.0	&	12		23		35	&	-123	&	7	&	20.61	&	1.50	&	\nodata		 	&	727	&	\nodata	&	\nodata	\\	
1240	&	204.9	&	551.5	&	12		30		41.2	&	12		19		46	&	-120	&	-222	&	20.23	&	1.76	&	\nodata		 	&	1359	&	\nodata	&	\nodata	\\	
1244	&	675.8	&	553.0	&	12		30		40.3	&	12		27		37	&	-133	&	249	&	19.90	&	1.23	&	\nodata		 	&	1913	&	\nodata	&	\nodata	\\	
1247	&	134.7	&	554.8	&	12		30		41.1	&	12		18		36	&	-122	&	-292	&	20.24	&	1.43	&	\nodata		 	&	1672	&	\nodata	&	\nodata	\\	
1251	&	437.7	&	556.4	&	12		30		40.5	&	12		23		39	&	-131	&	11	&	20.19	&	3.42	&	103	$\pm$	128	&	\nodata	&	\nodata	&	\nodata	\\	
1254	&	730.7	&	556.8	&	12		30		40.0	&	12		28		32	&	-138	&	304	&	21.12	&	1.37	&	896	$\pm$	102	&	877	&	\nodata	&	\nodata	\\	
1264	&	476.1	&	560.8	&	12		30		40.1	&	12		24		17	&	-136	&	49	&	19.29	&	2.62	&	\nodata		 	&	39	&	\nodata	&	\nodata	\\	
1265	&	 38.4	&	561.0	&	12		30		40.8	&	12		16		59	&	-126	&	-389	&	19.80	&	1.23	&	1045	$\pm$	86	&	\nodata	&	\nodata	&	1118	\\	
1280	&	 65.9	&	567.5	&	12		30		40.3	&	12		17		27	&	-133	&	-361	&	19.31	&	1.22	&	-107	$\pm$	73	&	0	&	\nodata	&	\nodata	\\	
1286	&	553.2	&	570.8	&	12		30		39.3	&	12		25		34	&	-148	&	126	&	20.86	&	1.24	&	971	$\pm$	158	&	\nodata	&	\nodata	&	\nodata	\\	
1290	&	249.0	&	572.2	&	12		30		39.7	&	12		20		30	&	-142	&	-178	&	20.77	&	1.50	&	\nodata		 	&	728	&	\nodata	&	\nodata	\\	
1291	&	167.0	&	573.2	&	12		30		39.8	&	12		19		08	&	-141	&	-260	&	19.46	&	1.96	&	\nodata		 	&	-57	&	\nodata	&	\nodata	\\	
1293	&	712.4	&	574.7	&	12		30		38.8	&	12		28		13	&	-156	&	285	&	20.13	&	1.45	&	889	$\pm$	61	&	776	&	\nodata	&	\nodata	\\	
1298	&	634.5	&	576.6	&	12		30		38.8	&	12		26		55	&	-156	&	207	&	19.70	&	1.10	&	\nodata		 	&	195	&	\nodata	&	\nodata	\\	
1301	&	311.7	&	578.6	&	12		30		39.2	&	12		21		32	&	-150	&	-116	&	20.03	&	1.33	&	\nodata		 	&	1053	&	\nodata	&	\nodata	\\	
1307	&	419.6	&	580.2	&	12		30		38.9	&	12		23		20	&	-154	&	-8	&	21.00	&	0.99	&	-118	$\pm$	129	&	\nodata	&	\nodata	&	\nodata	\\	
1309	&	448.7	&	581.3	&	12		30		38.8	&	12		23		49	&	-156	&	21	&	19.75	&	1.52	&	822	$\pm$	117	&	728	&	\nodata	&	\nodata	\\	
1313	&	688.8	&	583.1	&	12		30		38.2	&	12		27		50	&	-163	&	262	&	20.93	&	1.35	&	\nodata		 	&	1264	&	\nodata	&	\nodata	\\	
1316	&	303.5	&	584.7	&	12		30		38.8	&	12		21		24	&	-156	&	-124	&	20.20	&	1.72	&	1421	$\pm$	115	&	\nodata	&	\nodata	&	\nodata	\\	
1322	&	338.6	&	587.9	&	12		30		38.5	&	12		21		59	&	-160	&	-89	&	20.45	&	1.29	&	\nodata		 	&	1333	&	\nodata	&	\nodata	\\	
1336	&	209.8	&	591.3	&	12		30		38.5	&	12		19		50	&	-160	&	-218	&	19.87	&	1.48	&	\nodata		 	&	969	&	\nodata	&	\nodata	\\	
1341	&	523.8	&	592.9	&	12		30		37.8	&	12		25		04	&	-169	&	96	&	19.17	&	2.45	&	\nodata		 	&	-41	&	\nodata	&	\nodata	\\	
1344	&	427.2	&	594.0	&	12		30		37.9	&	12		23		27	&	-168	&	-1	&	19.75	&	1.49	&	993	$\pm$	159	&	982	&	\nodata	&	\nodata	\\	
1346	&	716.6	&	594.3	&	12		30		37.4	&	12		28		17	&	-175	&	289	&	19.57	&	0.72	&	\nodata		 	&	-178	&	\nodata	&	\nodata	\\	
1351	&	 60.4	&	597.3	&	12		30		38.3	&	12		17		20	&	-163	&	-368	&	19.79	&	1.56	&	\nodata		 	&	1705	&	\nodata	&	1783	\\	
1353	&	226.3	&	597.6	&	12		30		38.0	&	12		20		06	&	-167	&	-202	&	20.66	&	1.20	&	\nodata		 	&	1980	&	\nodata	&	\nodata	\\	
1354	&	477.5	&	598.8	&	12		30		37.5	&	12		24		18	&	-174	&	50	&	21.12	&	1.39	&	1438	$\pm$	84	&	\nodata	&	\nodata	&	\nodata	\\	
1356	&	361.5	&	599.4	&	12		30		37.7	&	12		22		22	&	-172	&	-66	&	18.91	&	3.02	&	\nodata		 	&	3	&	\nodata	&	\nodata	\\	
1367	&	584.9	&	606.8	&	12		30		36.8	&	12		26		05	&	-185	&	157	&	20.48	&	1.33	&	1221	$\pm$	94	&	1278	&	\nodata	&	\nodata	\\	
1370	&	177.9	&	607.4	&	12		30		37.4	&	12		19		18	&	-176	&	-250	&	19.21	&	1.60	&	1026	$\pm$	65	&	1041	&	\nodata	&	1337	\\	
1382	&	417.3	&	613.9	&	12		30		36.6	&	12		23		17	&	-188	&	-11	&	20.51	&	1.12	&	1492	$\pm$	107	&	1472	&	\nodata	&	\nodata	\\	
1391	&	228.9	&	618.8	&	12		30		36.6	&	12		20		08	&	-188	&	-200	&	19.61	&	1.47	&	\nodata		 	&	1186	&	\nodata	&	\nodata	\\	
1409	&	668.0	&	625.8	&	12		30		35.4	&	12		27		28	&	-206	&	240	&	20.17	&	1.28	&	\nodata		 	&	1103	&	\nodata	&	\nodata	\\	
1431	&	 84.0	&	640.7	&	12		30		35.3	&	12		17		43	&	-207	&	-345	&	20.72	&	1.09	&	\nodata		 	&	1303	&	\nodata	&	\nodata	\\	
1433	&	119.7	&	642.1	&	12		30		35.1	&	12		18		19	&	-209	&	-309	&	19.62	&	1.47	&	2046	$\pm$	109	&	1917	&	\nodata	&	1924	\\	
1434	&	618.6	&	642.3	&	12		30		34.3	&	12		26		38	&	-221	&	190	&	21.29	&	1.29	&	1445	$\pm$	108	&	\nodata	&	\nodata	&	\nodata	\\	
1449	&	387.3	&	650.4	&	12		30		34.1	&	12		22		46	&	-224	&	-42	&	20.92	&	1.24	&	\nodata		 	&	1067	&	\nodata	&	\nodata	\\	
1457	&	675.1	&	652.7	&	12		30		33.5	&	12		27		34	&	-233	&	246	&	20.58	&	1.79	&	\nodata		 	&	784	&	\nodata	&	\nodata	\\	
1461	&	640.1	&	654.2	&	12		30		33.5	&	12		26		59	&	-233	&	211	&	20.83	&	1.06	&	\nodata		 	&	693	&	\nodata	&	\nodata	\\	
1463	&	 83.6	&	655.5	&	12		30		34.3	&	12		17		42	&	-222	&	-346	&	20.44	&	1.22	&	\nodata		 	&	1871	&	\nodata	&	\nodata	\\	
1469	&	721.7	&	658.1	&	12		30		33.1	&	12		28		21	&	-239	&	293	&	20.71	&	1.69	&	\nodata		 	&	1101	&	\nodata	&	\nodata	\\	
1472	&	 99.0	&	659.3	&	12		30		34.0	&	12		17		57	&	-226	&	-331	&	20.55	&	0.81	&	321	$\pm$	141	&	\nodata	&	\nodata	&	\nodata	\\	
1479	&	746.4	&	668.4	&	12		30		32.3	&	12		28		45	&	-250	&	317	&	19.59	&	1.31	&	624	$\pm$	115	&	447	&	\nodata	&	\nodata	\\	
1481	&	455.1	&	670.4	&	12		30		32.7	&	12		23		54	&	-245	&	26	&	20.54	&	1.35	&	\nodata		 	&	1789	&	\nodata	&	\nodata	\\	
1483	&	 40.3	&	673.1	&	12		30		33.2	&	12		16		58	&	-238	&	-390	&	21.09	&	1.17	&	\nodata		 	&	1625	&	\nodata	&	\nodata	\\	
1489	&	225.8	&	675.2	&	12		30		32.7	&	12		20		04	&	-245	&	-204	&	19.24	&	4.06	&	-58	$\pm$	172	&	\nodata	&	\nodata	&	\nodata	\\	
1490	&	638.5	&	675.6	&	12		30		32.0	&	12		26		57	&	-255	&	209	&	20.95	&	1.10	&	\nodata		 	&	1403	&	\nodata	&	\nodata	\\	
1497	&	 23.1	&	678.7	&	12		30		32.8	&	12		16		41	&	-243	&	-407	&	19.67	&	0.80	&	103	$\pm$	79	&	3	&	\nodata	&	\nodata	\\	
1504	&	583.9	&	683.6	&	12		30		31.6	&	12		26		02	&	-261	&	154	&	19.60	&	1.19	&	\nodata		 	&	626	&	\nodata	&	1056	\\	
1508	&	546.4	&	686.3	&	12		30		31.4	&	12		25		24	&	-263	&	116	&	21.39	&	1.19	&	2419	$\pm$	140	&	\nodata	&	\nodata	&	\nodata	\\	
1512	&	260.9	&	687.6	&	12		30		31.8	&	12		20		39	&	-258	&	-169	&	20.44	&	1.28	&	984	$\pm$	159	&	\nodata	&	\nodata	&	\nodata	\\	
1514	&	387.1	&	688.3	&	12		30		31.6	&	12		22		45	&	-261	&	-43	&	19.49	&	1.79	&	\nodata		 	&	1165	&	\nodata	&	\nodata	\\	
1530	&	676.5	&	695.2	&	12		30		30.6	&	12		27		34	&	-275	&	246	&	20.59	&	1.16	&	1688	$\pm$	99	&	\nodata	&	\nodata	&	1465	\\	
1531	&	108.1	&	695.4	&	12		30		31.5	&	12		18		06	&	-262	&	-322	&	20.67	&	0.69	&	\nodata		 	&	209	&	\nodata	&	\nodata	\\	
1538	&	397.9	&	701.3	&	12		30		30.6	&	12		22		56	&	-275	&	-32	&	19.23	&	1.66	&	943	$\pm$	84	&	1237	&	\nodata	&	993	\\	
1540	&	698.4	&	703.1	&	12		30		30.0	&	12		27		56	&	-284	&	268	&	20.17	&	1.52	&	665	$\pm$	94	&	587	&	\nodata	&	\nodata	\\	
1546	&	497.2	&	706.1	&	12		30		30.2	&	12		24		35	&	-282	&	67	&	21.18	&	1.43	&	1452	$\pm$	92	&	\nodata	&	\nodata	&	\nodata	\\	
1548	&	107.6	&	707.7	&	12		30		30.7	&	12		18		05	&	-274	&	-323	&	19.63	&	1.55	&	\nodata		 	&	1735	&	\nodata	&	1680	\\	
1551	&	371.7	&	709.3	&	12		30		30.1	&	12		22		29	&	-282	&	-59	&	17.96	&	2.99	&	-105	$\pm$	88	&	not given	&	\nodata	&	\nodata	\\	
1563	&	639.8	&	716.3	&	12		30		29.2	&	12		26		57	&	-295	&	209	&	20.44	&	1.35	&	972	$\pm$	128	&	695	&	\nodata	&	\nodata	\\	
1565	&	470.5	&	717.8	&	12		30		29.4	&	12		24		08	&	-293	&	40	&	20.57	&	1.25	&	\nodata		 	&	1574	&	\nodata	&	\nodata	\\	
1575	&	500.2	&	726.3	&	12		30		28.8	&	12		24		37	&	-302	&	69	&	20.18	&	1.38	&	1161	$\pm$	47	&	\nodata	&	\nodata	&	\nodata	\\	
1577	&	364.0	&	727.6	&	12		30		28.9	&	12		22		21	&	-300	&	-67	&	20.70	&	1.30	&	\nodata		 	&	1366	&	\nodata	&	\nodata	\\	
1584	&	516.6	&	734.3	&	12		30		28.2	&	12		24		53	&	-311	&	85	&	19.20	&	1.58	&	\nodata		 	&	\nodata	&	\nodata	&	1340	\\	
1594	&	423.7	&	740.5	&	12		30		27.9	&	12		23		20	&	-315	&	-8	&	20.64	&	1.44	&	\nodata		 	&	1537	&	\nodata	&	\nodata	\\	
1602	&	 66.3	&	748.0	&	12		30		28.0	&	12		17		22	&	-314	&	-366	&	20.24	&	3.42	&	1009	$\pm$	135	&	\nodata	&	\nodata	&	\nodata	\\	
1611	&	580.8	&	752.7	&	12		30		26.8	&	12		25		57	&	-330	&	149	&	20.12	&	3.10	&	141	$\pm$	112	&	\nodata	&	\nodata	&	\nodata	\\	
1615	&	535.9	&	754.7	&	12		30		26.8	&	12		25		12	&	-331	&	104	&	19.44	&	1.46	&	\nodata		 	&	1168	&	\nodata	&	1405	\\	
1617	&	434.5	&	755.1	&	12		30		26.9	&	12		23		31	&	-329	&	3	&	19.18	&	1.38	&	1404	$\pm$	51	&	1369	&	\nodata	&	\nodata	\\	
1629	&	268.4	&	764.2	&	12		30		26.6	&	12		20		44	&	-335	&	-164	&	19.69	&	1.26	&	1026	$\pm$	131	&	\nodata	&	\nodata	&	\nodata	\\	
1631	&	604.7	&	767.8	&	12		30		25.8	&	12		26		21	&	-346	&	173	&	19.77	&	1.25	&	\nodata		 	&	1157	&	\nodata	&	1545	\\	
1634	&	479.1	&	769.2	&	12		30		25.9	&	12		24		15	&	-345	&	47	&	20.24	&	1.67	&	1587	$\pm$	61	&	\nodata	&	\nodata	&	\nodata	\\	
1644	&	594.3	&	773.8	&	12		30		25.4	&	12		26		10	&	-352	&	162	&	21.34	&	1.57	&	496	$\pm$	68	&	\nodata	&	\nodata	&	\nodata	\\	
1656	&	455.1	&	783.2	&	12		30		25.0	&	12		23		51	&	-358	&	23	&	17.06	&	3.20	&	26	$\pm$	126	&	\nodata	&	\nodata	&	\nodata	\\	
1657	&	700.5	&	783.7	&	12		30		24.5	&	12		27		56	&	-364	&	268	&	20.76	&	1.22	&	825	$\pm$	80	&	\nodata	&	\nodata	&	\nodata	\\	
1664	&	520.0	&	786.3	&	12		30		24.6	&	12		24		56	&	-363	&	88	&	19.49	&	1.51	&	\nodata		 	&	995	&	\nodata	&	\nodata	\\	
1668	&	153.4	&	792.1	&	12		30		24.8	&	12		18		49	&	-360	&	-279	&	18.79	&	3.14	&	18	$\pm$	121	&	\nodata	&	\nodata	&	\nodata	\\	
1676	&	603.7	&	797.8	&	12		30		23.7	&	12		26		19	&	-376	&	171	&	20.49	&	1.17	&	973	$\pm$	97	&	\nodata	&	\nodata	&	\nodata	\\	
1695	&	674.0	&	809.5	&	12		30		22.8	&	12		27		29	&	-389	&	241	&	19.61	&	1.64	&	1238	$\pm$	33	&	\nodata	&	\nodata	&	\nodata	\\	
1704	&	 73.0	&	815.8	&	12		30		23.4	&	12		17		28	&	-382	&	-360	&	18.22	&	2.32	&	-26	$\pm$	60	&	\nodata	&	\nodata	&	\nodata	\\	
1708	&	 10.5	&	823.7	&	12		30		22.9	&	12		16		25	&	-388	&	-423	&	19.00	&	0.66	&	-36	$\pm$	94	&	\nodata	&	\nodata	&	\nodata	\\	
1709	&	578.9	&	823.7	&	12		30		22.0	&	12		25		54	&	-401	&	146	&	20.50	&	1.39	&	\nodata		 	&	1790	&	\nodata	&	\nodata	\\	
1714	&	646.9	&	826.5	&	12		30		21.7	&	12		27		02	&	-406	&	214	&	19.95	&	2.90	&	23	$\pm$	81	&	\nodata	&	\nodata	&	\nodata	\\	
1716	&	488.8	&	827.7	&	12		30		21.9	&	12		24		23	&	-403	&	55	&	19.35	&	3.73	&	107	$\pm$	136	&	\nodata	&	\nodata	&	\nodata	\\	
1725	&	169.7	&	830.8	&	12		30		22.2	&	12		19		04	&	-399	&	-264	&	21.08	&	1.21	&	2064	$\pm$	117	&	\nodata	&	\nodata	&	\nodata	\\	
5001	&	386.2	&	403.7	&	12		30		51.0	&	12		22		51	&	23	&	-37	&	19.70	&	1.88	&	\nodata		 	&	680	&	\nodata	&	\nodata	\\	
5002	&	388.5	&	413.4	&	12		30		50.3	&	12		22		53	&	13	&	-35	&	21.39	&	0.06	&	\nodata		 	&	1252	&	\nodata	&	1276	\\	
5010	&	430.5	&	462.3	&	12		30		46.9	&	12		23		34	&	-36	&	6	&	20.47	&	1.61	&	\nodata		 	&	1401	&	\nodata	&	\nodata	\\	
5012	&	385.9	&	390.9	&	12		30		51.9	&	12		22		51	&	36	&	-37	&	20.57	&	1.96	&	\nodata		 	&	1718	&	\nodata	&	\nodata	\\	
5020	&	435.3	&	474.8	&	12		30		46.1	&	12		23		39	&	-49	&	11	&	20.57	&	1.43	&	\nodata		 	&	1450	&	\nodata	&	\nodata	\\	
5025	&	424.9	&	473.0	&	12		30		46.2	&	12		23		28	&	-47	&	0	&	20.34	&	1.43	&	\nodata		 	&	1762	&	\nodata	&	\nodata	\\	
5028	&	465.3	&	385.5	&	12		30		52.1	&	12		24		11	&	40	&	43	&	19.55	&	2	&	\nodata		 	&	1414	&	\nodata	&	\nodata	\\	
5053	&	389.7	&	427.2	&	12		30		49.4	&	12		22		54	&	0	&	-34	&	\nodata	&	\nodata	&	\nodata		 	&	1294	&	\nodata	&	\nodata	\\	
5055	&	399.8	&	416.5	&	12		30		50.1	&	12		23		04	&	10	&	-24	&	\nodata	&	\nodata	&	\nodata		 	&	1178	&	\nodata	&	\nodata	\\	
5058	&	409.4	&	399.4	&	12		30		51.2	&	12		23		15	&	27	&	-13	&	\nodata	&	\nodata	&	\nodata		 	&	195	&	\nodata	&	\nodata	\\	
5064	&	461.2	&	434.1	&	12		30		48.8	&	12		24		06	&	-9	&	38	&	\nodata	&	\nodata	&	\nodata		 	&	1278	&	\nodata	&	\nodata	\\	
5065	&	463.6	&	415.6	&	12		30		50.0	&	12		24		08	&	9	&	40	&	\nodata	&	\nodata	&	\nodata		 	&	1669	&	\nodata	&	1457	\\	
\enddata
\end{deluxetable}

\clearpage
\begin{deluxetable}{rrrccrrcccccc}
\tablenum{1}
\tablefontsize{\tiny}
\tablecolumns{13} 
\tablewidth{0pt}
\tablecaption{Positions, Photometry and Kinematics.}
\tablehead
{
   \colhead{ID}
 & \colhead{SX$^{"}$}
 & \colhead{SY$^{"}$}
 & \colhead{RA(2000)} 
 & \colhead{Dec(2000)} 
 & \colhead{$\Delta\alpha^{"}$}
 & \colhead{$\Delta\delta^{"}$} 
 & \colhead{T$_1$(mag)}
 & \colhead{C-T$_1$(mag)} 
 & \colhead{MOS (kms$^{-1}$)}
 & \colhead{K97 (kms$^{-1}$)}
 & \colhead{K00 (kms$^{-1}$)} 
 & \colhead{M90 (kms$^{-1}$)}
}
\startdata
5066	&	443.2	&	404.1	&	12		30		50.9	&	12		23		48	&	21	&	20	&	\nodata	&	\nodata	&	\nodata		 	&	1478	&	\nodata	&	\nodata	\\	
5067	&	434.4	&	400.1	&	12		30		51.2	&	12		23		40	&	26	&	12	&	\nodata	&	\nodata	&	\nodata		 	&	1173	&	\nodata	&	\nodata	\\	
5071	&	368.1	&	428.9	&	12		30		49.3	&	12		22		32	&	-2	&	-56	&	\nodata	&	\nodata	&	\nodata		 	&	1767	&	\nodata	&	\nodata	\\	
6003	&	 -3.6	&	132.6	&	12		31		10.1	&	12		16		28	&	304	&	-420	&	20.62	&	1.19	&	\nodata		 	&	\nodata	&	1741	&	\nodata	\\	
6004	&	-30.9	&	131.8	&	12		31		10.2	&	12		16		01	&	305	&	-447	&	20.92	&	1.10	&	\nodata		 	&	\nodata	&	1741	&	\nodata	\\	
7001	&	722.2	&	912.6	&	12		30		15.7	&	12		28		15	&	-494	&	287	&	18.66	&	0.68	&	52	$\pm$	111	&	\nodata	&	\nodata	&	\nodata	\\	
7002	&	109.6	&	906.8	&	12		30		17.1	&	12		18		02	&	-474	&	-326	&	19.67	&	0.54	&	-329	$\pm$	111	&	\nodata	&	\nodata	&	\nodata	\\	
7003	&	368.9	&	895.7	&	12		30		17.4	&	12		22		22	&	-469	&	-66	&	18.50	&	1.64	&	-310	$\pm$	181	&	\nodata	&	\nodata	&	\nodata	\\	
7004	&	458.2	&	893.2	&	12		30		17.4	&	12		23		51	&	-468	&	23	&	21.52	&	1.05	&	1786	$\pm$	174	&	\nodata	&	\nodata	&	\nodata	\\	
7005	&	  1.5	&	880.0	&	12		30		19.1	&	12		16		14	&	-444	&	-434	&	20.90	&	0.55	&	-16	$\pm$	105	&	\nodata	&	\nodata	&	\nodata	\\	
7006	&	530.3	&	865.8	&	12		30		19.2	&	12		25		04	&	-442	&	96	&	20.87	&	1.13	&	104	$\pm$	58	&	\nodata	&	\nodata	&	\nodata	\\	
7007	&	715.4	&	859.4	&	12		30		19.3	&	12		28		09	&	-440	&	281	&	21.16	&	1.18	&	138	$\pm$	94	&	\nodata	&	\nodata	&	\nodata	\\	
7008	&	232.9	&	867.2	&	12		30		19.6	&	12		20		06	&	-437	&	-202	&	21.67	&	2.33	&	2227	$\pm$	142	&	\nodata	&	\nodata	&	\nodata	\\	
7009	&	719.0	&	850.6	&	12		30		19.9	&	12		28		13	&	-432	&	285	&	19.84	&	1.93	&	1492	$\pm$	78	&	\nodata	&	\nodata	&	\nodata	\\	
7010	&	481.2	&	841.5	&	12		30		20.9	&	12		24		15	&	-417	&	47	&	17.32	&	3.14	&	133	$\pm$	71	&	\nodata	&	\nodata	&	\nodata	\\	
7011	&	874.1	&	829.4	&	12		30		21.1	&	12		30		49	&	-414	&	441	&	20.65	&	1.27	&	964	$\pm$	179	&	\nodata	&	\nodata	&	\nodata	\\	
7012	&	412.0	&	839.9	&	12		30		21.2	&	12		23		06	&	-414	&	-22	&	21.89	&	0.92	&	1205	$\pm$	141	&	\nodata	&	\nodata	&	\nodata	\\	
7013	&	852.5	&	810.9	&	12		30		22.4	&	12		30		28	&	-395	&	420	&	20.77	&	1.36	&	1004	$\pm$	101	&	\nodata	&	\nodata	&	\nodata	\\	
7014	&	907.1	&	781.4	&	12		30		24.3	&	12		31		23	&	-367	&	475	&	19.47	&	2.44	&	160	$\pm$	90	&	\nodata	&	\nodata	&	\nodata	\\	
7015	&	884.9	&	771.4	&	12		30		25.1	&	12		31		01	&	-356	&	453	&	19.78	&	1.14	&	-240	$\pm$	108	&	\nodata	&	\nodata	&	\nodata	\\	
7016	&	-25.3	&	719.4	&	12		30		30.1	&	12		15		52	&	-283	&	-456	&	20.99	&	1.21	&	1902	$\pm$	100	&	\nodata	&	\nodata	&	\nodata	\\	
7017	&	798.8	&	644.2	&	12		30		33.9	&	12		29		38	&	-227	&	370	&	20.06	&	2.17	&	94	$\pm$	101	&	\nodata	&	\nodata	&	\nodata	\\	
7018	&	 -7.1	&	621.7	&	12		30		36.7	&	12		16		12	&	-186	&	-436	&	18.91	&	1.69	&	-8	$\pm$	43	&	\nodata	&	\nodata	&	\nodata	\\	
7019	&	492.9	&	488.4	&	12		30		45.0	&	12		24		36	&	-64	&	68	&	20.30	&	2.03	&	1589	$\pm$	153	&	\nodata	&	\nodata	&	\nodata	\\	
7020	&	412.9	&	480.4	&	12		30		45.7	&	12		23		16	&	-54	&	-12	&	19.54	&	1.80	&	1231	$\pm$	89	&	\nodata	&	\nodata	&	\nodata	\\	
7021	&	-52.5	&	475.2	&	12		30		46.8	&	12		15		30	&	-38	&	-478	&	19.63	&	1.41	&	1225	$\pm$	88	&	\nodata	&	\nodata	&	\nodata	\\	
7022	&	 -1.4	&	463.8	&	12		30		47.5	&	12		16		22	&	-28	&	-426	&	21.41	&	1.09	&	-373	$\pm$	184	&	\nodata	&	\nodata	&	\nodata	\\	
7023	&	-51.1	&	458.7	&	12		30		47.9	&	12		15		32	&	-21	&	-476	&	21.04	&	1.70	&	415	$\pm$	134	&	\nodata	&	\nodata	&	\nodata	\\	
7024	&	475.6	&	427.0	&	12		30		49.2	&	12		24		20	&	-2	&	52	&	19.94	&	1.17	&	\nodata		 	&	\nodata	&	\nodata	&	1632	\\	
7025	&	-39.3	&	 -2.8	&	12		31		19.4	&	12		15		56	&	440	&	-452	&	19.94	&	1.69	&	1463	$\pm$	69	&	\nodata	&	\nodata	&	\nodata	\\	
7026	&	391.0	&	-16.2	&	12		31		19.6	&	12		23		06	&	443	&	-22	&	20.22	&	0.84	&	2	$\pm$	110	&	\nodata	&	\nodata	&	\nodata	\\	
7027	&	179.1	&	-26.0	&	12		31		20.7	&	12		19		35	&	458	&	-233	&	19.70	&	1.42	&	72	$\pm$	61	&	\nodata	&	\nodata	&	\nodata	\\	
7028	&	180.0	&	-48.5	&	12		31		22.2	&	12		19		36	&	481	&	-232	&	20.91	&	1.21	&	1345	$\pm$	91	&	\nodata	&	\nodata	&	\nodata	\\	
8001	&	 87.4	&	224.8	&	12		31		03.7	&	12		17		57	&	209	&	-331	&	18.70	&	2.93	&	159	$\pm$	127	&	not given	&	\nodata	&	\nodata	\\	
8002	&	203.2	&	452.9	&	12		30		47.9	&	12		19		47	&	-22	&	-221	&	20.52	&	1.39	&	1565	$\pm$	144	&	not given	&	\nodata	&	\nodata	\\	
8003	&	216.4	&	248.1	&	12		31	  	01.9	&	12		20		05	&	183	&	-203	&	20.87	&	1.24	&	1355	$\pm$	122	&	\nodata	&	\nodata	&	\nodata	\\	
8004	&	475.6	&	480.5	&	12		30		45.6	&	12		24		19	&	-56	&	51	&	20.10	&	1.51	&	1658	$\pm$	139	&	\nodata	&	\nodata	&	\nodata	\\	
8005	&	478.4	&	471.6	&	12		30		46.2	&	12		24		22	&	-47	&	54	&	19.80	&	1.34	&	\nodata		 	&	1934	&	\nodata	&	\nodata	\\	
8006	&	478.4	&	464.7	&	12		30		46.7	&	12		24		22	&	-40	&	54	&	19.80	&	1.35	&	1032	$\pm$	107	&	1109	&	\nodata	&	\nodata	\\	
8007	&	772.6	&	531.3	&	12		30		41.6	&	12		29		15	&	-114	&	347	&	19.46	&	1.58	&	1183	$\pm$	32	&	1106	&	\nodata	&	\nodata	\\	
8008	&	794.8	&	592.8	&	12		30		37.4	&	12		29		35	&	-176	&	367	&	20.36	&	1.37	&	\nodata		 	&	1296	&	\nodata	&	\nodata	\\	
8051	&	439.6	&	622.6	&	12		30		36.0	&	12		23		39	&	-197	&	11	&	20.76	&	1.35	&	\nodata		 	&	915	&	\nodata	&	\nodata	\\	
8052	&	489.8	&	407.4	&	12		30		50.6	&	12		24		35	&	17	&	67	&	\nodata	&	\nodata	&	\nodata		 	&	1477	&	\nodata	&	\nodata	\\	
8053	&	646.7	&	251.3	&	12		31		01.0	&	12		27		16	&	169	&	228	&	\nodata	&	\nodata	&	\nodata		 	&	15	&	\nodata	&	\nodata	\\	
8054	&	478.8	&	300.2	&	12		30		57.9	&	12		24		26	&	125	&	58	&	19.95	&	1.73	&	\nodata		 	&	1169	&	\nodata	&	\nodata	\\	
8055	&	255.5	&	363.8	&	12		30		53.9	&	12		20		41	&	66	&	-167	&	20.59	&	1.58	&	\nodata		 	&	467	&	\nodata	&	\nodata	\\	
8056	&	425.9	&	495.3	&	12		30		44.7	&	12		23		29	&	-69	&	1	&	20.04	&	1.78	&	\nodata		 	&	1368	&	\nodata	&	\nodata	\\	
9001	&	480.5	&	636.7	&	12		30		34.9	&	12		24		20	&	-212	&	52	&	\nodata	&	\nodata	&	\nodata		 	&	1338	&	\nodata	&	\nodata	\\	
9002	&	545.7	&	495.1	&	12		30		44.5	&	12		25		29	&	-72	&	121	&	\nodata	&	\nodata	&	\nodata		 	&	1100	&	\nodata	&	\nodata	\\	
9051	&	363.7	&	437.2	&	12		30		48.7	&	12		22		28	&	-10	&	-60	&	\nodata	&	\nodata	&	\nodata		 	&	1879	&	\nodata	&	\nodata	\\	
9052	&	320.5	&	427.7	&	12		30		49.5	&	12		21		45	&	1	&	-103	&	\nodata	&	\nodata	&	1763	$\pm$	127	&	1480	&	\nodata	&	\nodata	\\	
9053	&	796.0	&	433.1	&	12		30		48.3	&	12		29		41	&	-16	&	373	&	\nodata	&	\nodata	&	\nodata		 	&	796	&	\nodata	&	\nodata	\\	
\enddata
\end{deluxetable}

\clearpage
\begin{deluxetable}{ccl}
\tablenum{2}
\tablefontsize{\small}
\tablecaption{Revised Identification Numbers.}
\tablewidth{0pt}
\tablehead{
\colhead{Old ID} & \colhead{New ID} & \colhead{Comments}
}
\startdata
354  &	8001	& Double, unresolved in S81; unique MOS target identified\\
408  &      8003  & Double, unresolved in S81; unique MOS target identified\\
410  &      8053  & Double, unresolved in S81; Keck target; both components blue\\
526  &	8054  & Double, unresolved in S81; Keck target; both components red\\
715  &      8055  & Very close double, unresolved in S81; Keck target; both components red\\
827  &	8052  & Double, unresolved in S81; Keck target; both components red\\
868  &      9052  & No reliable photometry; cosmic ray in photometric database?\\
881  &      9053  & No reliable photometry\\
892  &      9051  & No entry in photometry database; clearly seen in M87 direct frames\\
944  &      8002  & Double, unresolved in S81; unique MOS target identified\\
978  &      8006  & Multiple, unresolved in S81; reported twice in K97; unique MOS target identified\\
1013 &	8004  & Double, unresolved in S81; unique MOS target identified\\
1065 &	9002  & Double, unresolved in S81; Keck target; one component blue, one red\\
1067 &	8056  & Very close double, unresolved in S81; Keck target; both components red\\
1180 &      8007  & Double, unresolved in S81; unique MOS target identified\\
1340 &	8008  & Double, unresolved in S81; unique MOS target identified\\
1400 &      8051  & Very close double, unresolved in S81; Keck target; both components blue\\
1425 &	9001  & Double, unresolved in S81; Keck target; one component blue, one red\\
5003 &      5053  & No photometry\\
5005 &	5055  & No photometry\\
5008 &	5058  & No photometry\\
5014 &	5064	& No photometry\\
5015 &	5065  & No photometry\\
5016 &	5066	& No photometry\\
5017 &	5067  & No photometry\\
5021 &      5071  & No photometry\\
5024 &  8006 & Recognized as redundant listing of 978 in K97; see entry above\\
5026 & 8005  & Multiple, part of unresolved target 978 in S81; given twice in K97; merged\\
\enddata
\end{deluxetable}

\begin{deluxetable}{ll}
\tablenum{3}
\tablecaption{Properties of the MOS Detector and Grism.}
\tablewidth{0pt}
\tablehead{
\colhead{Item} & \colhead{Attribute}
}
\startdata
Detector & $2048$x$2048$ STIS2 CCD \\
Image Scale & $0\uparcs44$ per $21\mu$ pixel\\
QE & $0.85$ at $5200\AA$\\
Read Noise & $\sim9$ electrons\\
Field of View & $\sim9.5$ arcmin\\
Grism & B400\\
Zero Deviation& $5186\AA$\\
Dispersion & $3.59\AA$ per pixel\\
Filter central wavelength & $5100\AA$\\
Filter bandwidth & $1200\AA$\\
\enddata
\end{deluxetable}

\clearpage
\begin{deluxetable}{ccc}
\tablenum{4}
\tablecaption{Galactic Globular Clusters Used as Templates.}
\tablewidth{0pt}
\tablehead{
\colhead{Cluster} & \colhead{[Fe/H]\tablenotemark{a} (dex)} & \colhead{Heliocentric 
velocity\tablenotemark{a} (km/s)}
}
\startdata
NGC$6171$ &	$-1.04$	&	$-33.6$\\
NGC$6205$ &	$-1.54$ &	$-246.6$\\
NGC$6356$ &	$-0.50$	&	$27.0$\\
NGC$6402$ &	$-1.39$ &	$-66.1$\\
NGC$6528$ &  $-0.17$ &	$184.9$\\
NGC$6624$ &  $-0.42$ &	$53.9$\\
\enddata
\tablenotetext{a}{ Harris (1996).}
\end{deluxetable}

\begin{deluxetable}{cccl}
\tablenum{5}
\tablecaption{Spectroscopic Frames of the M87 GCS.}
\tablewidth{0pt}
\tablehead{
\colhead{Field} & \colhead{Exposure time} 
& \colhead{Airmass} &\colhead{Comments}
}
\startdata

Central 1  & 2 x 3600s (Night 2) & $1.03-1.12$ & Scattered light; 
some cloud.\\

Central 2  & 2 x 3600s (Night 2) & $1.13-1.88$
& Scattered light; some cloud.\\

Central 2& 1 x 3600s (Night 3) & $1.03-1.02$ & Scattered light.\\

Southeast & 2 x 3600s (Night 3) & $1.02-1.25$ 
& Scattered light, humid, poor seeing.\\

Southeast & 2 x 3600s (Night 4) & $1.14-1.85$ &Scattered light eliminated\\

Northwest & 2 x 3600s (Night 4) & $1.01-1.10$ 
& Mask  
slightly rotated; about $50\%$ of targets on slitlets.\\

Northwest & 2 x 3600s (Night 5) & $1.01-1.15$&New NW mask cut and mounted; no problems.\\ 

Southwest & 2 x 3600s (Night 5) & $1.21-2.10$ &No problems.\\
\enddata
\end{deluxetable}


\begin{thebibliography}{}
\bibitem[Ashman and Zepf(1992)]{ash92} Ashman, K.M. and Zepf, S.E. 1992, \apj, 384, 50
\bibitem[Cohen(2000)]{coh00} Cohen, J.G. 2000, \aj, 119, 162
\bibitem[Cohen, Blakeslee and Ryzhov(1998)]{coh98} Cohen, J.G., Blakeslee, J.P.
	and Ryzhov, A. 1998, \apj, 496, 808 
\bibitem[Cohen and Ryzhov(1997)]{coh97} Cohen, J.G. and Ryzhov, A. 1997, \apj, 486, 230
\bibitem[C\^ot\'e, Marzke and West(1998)]{cot98} C\^ot\'e, P., Marzke, R.O. and
	West, M.J. 1998, \apj, 501, 554
\bibitem[C\^ot\'e et al.(2001)]{cot01} C\^ot\'e, P., McLaughlin, D.E., Hanes, D.A.,
        Bridges, T.J., Geisler, D., Merritt, D., Hesser, J.E., and Harris, G.L.H.
        2001, \apj, to be submitted.
\bibitem[Eggen, Lynden-Bell and Sandage(1962)]{egg62} Eggen, O.J., Lynden-Bell, D.
	and Sandage, A.R. 1962, \apj, 136, 748
\bibitem[Elson and Santiago(1996)]{els96} Elson, R.A.W. and Santiago, B.X.
	1996, \mnras, 278, 617
\bibitem[Geisler(1996)]{gei96} Geisler, D. 1996, \aj, 111, 480.
\bibitem[Geisler, Lee and Kim(2001)]{gei01} Geisler, D., Lee, M.G. and Kim, S.C.  2001,
	{\em in preparation}
\bibitem[Hanes(1977)]{han77} Hanes, D.A. 1977, Mem.R.A.S. 84, 45
\bibitem[Harris(1986)]{har86} Harris, W.E. 1986, \aj, 91, 822
\bibitem[Harris(1996)]{har96} Harris, W.E. 1996, \aj, 112, 1487
\bibitem[Harris, Harris and McLaughlin(1998)]{har98} Harris, W.E., Harris, G.L.H. and 
	McLaughlin, D.E. 1998, \aj, 115, 1801
\bibitem[Huchra and Brodie(1987)]{huc87} Huchra, J.P. and Brodie, J.P. 1987,
	\aj, 93, 779
\bibitem[Kissler-Patig and Gebhardt(1998)]{kis98} Kissler-Patig, M. and Gebhardt, K. 1998,
	\aj, 116, 2237
\bibitem[Kundu et al.(1999)]{kun99}Kundu, A., Whitmore, B.C., Sparks, W.B.,
        Macchetto, F.D., Zepf, S.E. and Ashman, K.M. 1999, \apj, 513, 733
\bibitem[Le Fevre et al.(1994)]{lef94}Le Fevre, O., Crampton, D., Felenbok, P.,
	and Monnet, G. 1994, \aap, 282, 325
\bibitem[McLaughlin, Harris and Hanes(1994)]{mcl94} McLaughlin, D.E.,
	Harris, W.E. and Hanes, D.A. 1994, \apj, 422, 486
\bibitem[Mould, Oke and Nemec(1987)]{mou87} Mould, J.R., Oke, J.B.,
	and Nemec, J.M. 1987, \aj, 93, 53 
\bibitem[Mould et al.(1990)]{mou90} Mould, J.R., Oke, J.B., de Zeeuw, P.T.
	and Nemec, J.M. 1990, \aj, 99, 1823
\bibitem[Strom et al.(1981)]{str81} Strom, S.E., Forte, J.C., Harris, W.E.,
	Strom, K.M., Wells, D.C. and Smith, M.G. 1981, \apj, 245, 416
\bibitem[Whitmore et al.(1995)]{whi95} Whitmore, B.C., Sparks, W.B., Lucas, R.A.,
	Macchetto, F.Duccio, and Biretta, J.A. 1995, \apjl, 454L, 73



\end{thebibliography}
\end{document}